\documentclass{aa}

\usepackage[varg]{txfonts}
\usepackage{graphicx}
\usepackage{algorithm}
\usepackage{algorithmic}
\usepackage{color}
\usepackage{placeins}

\usepackage{CJK}

\usepackage{natbib,twoopt}
\usepackage[breaklinks=true]{hyperref} 
\bibpunct{(}{)}{;}{a}{}{,} 

\authorrunning{Lin et al.}
\titlerunning{CLAP. I. Resolving miscalibration for photo-$z$ estimation}

\newcommand{\Treyer}{Treyer et al. (in prep.)}

\begin{document}
\begin{CJK*}{UTF8}{gbsn}

\title{CLAP. I. Resolving miscalibration for deep learning-based galaxy photometric redshift estimation}

\author{Qiufan Lin (林秋帆)\inst{\ref{inst1}}\thanks{E-mail: linqf@pcl.ac.cn}
\and Hengxin Ruan (阮恒心)\inst{\ref{inst1}}
\and Dominique Fouchez\inst{\ref{inst2}}
\and Shupei Chen (陈树沛)\inst{\ref{inst1}}
\and Rui Li (李瑞)\inst{\ref{inst3}}
\and Paulo Montero-Camacho\inst{\ref{inst1}}
\and Nicola R. Napolitano\inst{\ref{inst4}, \ref{inst5}, \ref{inst6}}
\and Yuan-Sen Ting (丁源森)\inst{\ref{inst7}, \ref{inst8}, \ref{inst9}, \ref{inst10}}
\and Wei Zhang (张伟)\inst{\ref{inst1}}}

\institute{Pengcheng Laboratory, Nanshan District, Shenzhen, Guangdong 518000, P.R. China\label{inst1}
\and Aix-Marseille Universit\'{e}, CNRS/IN2P3, CPPM, Marseille 13009, France\label{inst2}
\and School of Physics and Microelectronics, Zhengzhou University, Zhengzhou, Henan 450001, P.R. China\label{inst3}
\and Department of Physics E. Pancini, University Federico II, Via Cinthia 6, I-80126, Naples, Italy\label{inst4}
\and School of Physics and Astronomy, Sun Yat-sen University, Zhuhai Campus, 2 Daxue Road, Tangjia, Zhuhai, Guangdong 519082, P.R.China\label{inst5}
\and CSST Science Center for Guangdong-Hong Kong-Macau Great Bay Area, Zhuhai, Guangdong 519082, P.R. China\label{inst6}
\and Department of Astronomy, The Ohio State University, Columbus, OH 43210, USA\label{inst7}
\and Center for Cosmology and AstroParticle Physics (CCAPP), The Ohio State University, Columbus, OH 43210, USA\label{inst8}
\and Research School of Astronomy \& Astrophysics, Australian National University, Cotter Rd., Weston, ACT 2611, Australia\label{inst9}
\and School of Computing, Australian National University, Acton, ACT 2601, Australia\label{inst10}
}

\date{Received; accepted}

\abstract{Obtaining well-calibrated photometric redshift probability densities for galaxies without a spectroscopic measurement remains a challenge. Deep learning discriminative models, typically fed with multi-band galaxy images, can produce outputs that mimic probability densities and achieve state-of-the-art accuracy. However, several previous studies have found that such models may be affected by miscalibration, an issue that would result in discrepancies between the model outputs and the actual distributions of true redshifts. Our work develops a novel method called the \textbf{C}ontrastive \textbf{L}earning and \textbf{A}daptive KNN for \textbf{P}hotometric Redshift (CLAP) that resolves this issue. It leverages supervised contrastive learning (SCL) and $k$-nearest neighbours (KNN) to construct and calibrate raw probability density estimates, and implements a refitting procedure to resume end-to-end discriminative models ready to produce final estimates for large-scale imaging data, bypassing the intensive computation required for KNN. The harmonic mean is adopted to combine an ensemble of estimates from multiple realisations for improving accuracy. Our experiments demonstrate that CLAP takes advantage of both deep learning and KNN, outperforming benchmark methods on the calibration of probability density estimates and retaining high accuracy and computational efficiency. With reference to CLAP, a deeper investigation on miscalibration for conventional deep learning is presented. We point out that miscalibration is particularly sensitive to the method-induced excessive correlations among data instances in addition to the unaccounted-for epistemic uncertainties. Reducing the uncertainties may not guarantee the removal of miscalibration due to the presence of such excessive correlations, yet this is a problem for conventional methods rather than CLAP. These discussions underscore the robustness of CLAP for obtaining photometric redshift probability densities required by astrophysical and cosmological applications. This is the first paper in our series on CLAP.
}


\maketitle

\section{Introduction} \label{sec:intro}

In cosmological and extragalactic studies, the cosmological redshift $z$ characterises the cosmic distance, and it is a crucial quantity for probing the properties of galaxies and the evolution of the Universe. The most direct and accurate way to measure redshift is through spectroscopy. However, spectroscopic redshifts (spec-$z$) are highly time intensive to obtain. There have been multiple ongoing or planned galaxy imaging surveys in recent years, such as the Kilo-Degree Survey \citep[KiDS;][]{deJong2013}, the Dark Energy Survey \citep[DES;][]{DES2016}, the \textit{Euclid} survey \citep{Laureijs2011}, the Hyper Suprime-Cam \citep[HSC;][]{Aihara2018}, the \textit{Nancy Grace Roman} Space Telescope \citep{Spergel2015}, the \textit{Vera C. Rubin} Observatory Legacy Survey of Space and Time \citep[LSST;][]{Ivezic2019}, and the China Space Station Telescope \citep[CSST;][]{Zhan2018}. They require redshift estimates for hundreds of millions or billions of galaxies for which spectroscopic measurements are infeasible. Given such extremely rich datasets, photometric redshifts (photo-$z$), measured typically using photometry, have become an alternative in order to meet the needs for large imaging surveys.

The idea behind photometric redshift estimation for individual galaxies lies in the mapping between the observed galaxy photometric or morphological properties and the redshift. Broadly speaking, two categories of methods are commonly leveraged for deriving individual redshift estimates (see \citet{Salvato2019} and \citet{NG2022}). Template-fitting methods \citep[e.g.][]{Arnouts1999, BPZ2000, Feldmann2006, Ilbert2006, Greisel2015, Leistedt2019} determine redshifts by finding the best fit between the galaxy spectral energy distribution (SED) and a library of SED templates covering different physical and morphological properties such as galaxy types. Data driven or empirical methods determine redshifts by empirically learning the mapping from photometric data to redshift estimates without any theoretical modelling. These include unsupervised learning approaches in which known redshifts are given as external information, such as $k$-nearest neighbours \citep[KNN; e.g.][]{Zhang2013, Beck2016, DeVicente2016, Speagle2019, Han2021, Luken2022}, self-organising maps \citep[SOMs; e.g.][]{Way2012, Carrasco2014, Speagle2017, Buchs2019, Wilson2020}, and supervised learning approaches in which known redshifts are used as labels, such as artificial neural networks \citep[ANNs; e.g.][]{Collister2004, Brescia2014, Bonnett2015, Cavuoti2015, Hoyle2016, Sadeh2016, Cavuoti2017, Bilicki2018, DP2018, Pasquet2019, Mu2020photoz, Schuldt2021, Dey2022, AO2023, Treyer2024}, decision trees and random forest \citep[e.g.][]{Carliles2010, CB2013, Luken2022}, boosted decision trees \citep[e.g.][]{Gerdes2010}, support vector machines \citep[SVMs; e.g.][]{Jones2017}, and Gaussian mixture models \citep[GMMs; e.g.][]{DP2018, JH2019, Hatfield2020, Ansari2021}.

Thanks to the recent advances in deep learning, a few studies \citep[e.g.][]{DP2018, Pasquet2019, Schuldt2021, Dey2022, AO2023, Treyer2024} have developed deep neural networks (DNNs) to predict photometric redshifts and achieve state-of-the-art predictive accuracy. These discriminative models, developed in a supervised end-to-end manner and usually having a large number of trainable weights, directly take multi-band stamp images of galaxies as inputs and produce redshift estimates. They are trained iteratively with mini-batches of training data, minimising a loss function via gradient descent in which spectroscopic redshifts are used as labels. In this way, the models can automatically establish a mapping from input images to the target redshifts. Studies such as \citet{Henghes2022}, \citet{Li2022GaZNets}, \citet{Zhou2022}, and \citet{Zhou2022b} explored hybrid networks that are fed with both images and multi-band photometry or fluxes. The outperformance of such image-based methods in accuracy over photometry-only methods suggests that other than photometry there may be extra information such as galaxy morphology encoded in images that helps improve redshift estimation \citep{Soo2018}.

Photometric redshift estimates for individual galaxies are usually considered in two forms: (a) the point estimate $z_{photo}$ derived from photometric data $d$ (i.e. photometry or images) and (b) the (posterior) probability density estimate $p(z|d)$ over possible values given photometric data $d$, accounting for imperfect knowledge in determining redshifts. This work concentrates on probability density estimation via deep learning. 

Obtaining well-calibrated redshift probability density estimates is a challenging task, as reported for a large array of redshift estimation methods \citep[e.g.][]{Wittman2016, Tanaka2018, Amaro2019, Desprez2020, Schmidt2020, Kodra2023}. There have been several techniques coupled with deep learning models used to express redshift probability densities, including the softmax function \citep{Pasquet2019, Treyer2024}, GMMs \citep{DP2018}, and Bayesian neural networks \citep[BNNs;][]{Zhou2022b}, outputting vectors covering a redshift range and normalised to unity. Despite their high accuracy, the calibration of such estimates produced by neural networks remains an open issue. From a statistical viewpoint, a well-calibrated probability density estimate should fulfil the frequentist definition: for a given sample of galaxies, the integrated probability within any arbitrary redshift range must describe the relative frequency of true redshifts falling in this range \citep{Dey2021, Dey2022b}. While there are proofs that ANNs are capable of providing reliable posterior probability estimates \citep{Richard1991, Rojas1996}, several studies have shown evidence that the probability density outputs from discriminative models especially those with high capacity may suffer from miscalibration, that is, a model's confidence is not consistent with its accuracy \citep[e.g.][]{Guo2017, Thulasidasan2019, Minderer2021, Wen2021}. In other words, miscalibration would lead to a deviation between the estimated $p'(z|d)$ and the empirically correct $p(z|d)$. \citet{Guo2017} pointed out that the extent of miscalibration generally grows with the model capacity for typical neural networks. As we show, neural networks similar to the one developed by \citet{Treyer2024} (with $\sim$60 layers and $\sim$7 million trainable parameters) already exhibit clear miscalibration behaviours. Furthermore, the variability stemming from the iterative training procedure would contribute additional uncertainties and increase the potential risk of miscalibration \citep{Huang2021, Huang2023}.

Unlike point estimates that only give the collapsed values, well-calibrated probability density estimates are capable of describing the intrinsic dispersions due to the complex many-to-many mapping from photometric data to redshift, and describing the uncertainties due to the method implemented for redshift estimation and errors in photometric data. They are thus preferred for precision cosmology, in which one needs to understand and incorporate the full uncertainties in photometric redshift estimation into analysis \citep[e.g.][]{Mandelbaum2008, Myers2009, Bonnett2016, AH2019, RuizZapatero2023, Zhang2023}. On the other hand, this leads to a strong requirement for the calibration of probability density estimates. The potential miscalibration issue associated with conventional deep learning would result in unreliable characterisation of uncertainties and would bias the photometric redshift estimation, consequently degrading the cosmological inference. In particular, miscalibrated probability densities for individual galaxies may induce biases on the mean redshifts estimated in tomographic bins and lead to poorly reconstructed redshift distribution over a galaxy sample, which are severe for weak lensing tomography \citep[e.g.][]{Ma2006, Huterer2006, Laureijs2011, Joudaki2020, Hildebrandt2020, Ilbert2021}.

In our previous work \citep{Lin2022} we developed a set of consecutive steps to correct biases in photometric redshift estimation with deep learning, potentially overcoming the impact of miscalibration on the mean redshifts in tomographic bins. However, this approach is limited to point estimate analysis, and suffers from a trade-off between constraining estimation errors and correcting biases, which originates from retraining a fraction of a network using a subset of training data and soft labels that enlarges estimation errors. We thus attempt to find a better solution to tackle the miscalibration issue for probability density estimation.

In this work we propose the \textbf{C}ontrastive \textbf{L}earning and \textbf{A}daptive KNN for \textbf{P}hotometric Redshift (CLAP), a novel method that resolves the miscalibration encountered by conventional deep learning in the context of photometric redshift probability density estimation. CLAP leverages the gains from KNN and retains the advantages of deep learning simultaneously. It turns a discriminative model into a contrastive learning framework that projects galaxy images and additional input data to redshift-sensitive latent vectors, followed by an adaptive KNN algorithm and a KNN-enabled recalibration procedure to produce locally calibrated probability density estimates, which leverage diagnostics with the probability integral transform \citep[PIT;][]{Gass2001}. The contrastive learning is coupled with supervised learning in which spectroscopic redshifts are used as labels. We give a solution for alleviating the reliance on the expensive computation required for KNN, and suggest a proper way to combine an ensemble of probability density estimates for reducing uncertainties. We demonstrate CLAP using data separately from the Sloan Digital Sky Survey \citep[SDSS;][]{Eisenstein2011}, the Canada-France-Hawaii Telescope Legacy Survey \citep[CFHTLS;][]{Gwyn2012}, and the Kilo-Degree Survey \citep[KiDS;][]{deJong2013} As we show, the main advantages of CLAP include the following:
\begin{itemize}
    \item a substantial improvement on the calibration of probability density estimates thanks to KNN;
    \item a high accuracy comparable to that obtained by conventional image-based deep learning methods, better than photometry-only KNN approaches such as \citet{Beck2016};
    \item computational efficiency of end-to-end deep learning models retained by bypassing the intensive computation required for KNN.
\end{itemize}
This is in line with other works that demonstrate the merits of combining deep learning with KNN \citep[e.g.][]{PM2018, Chen2020, Dwibedi2021, Sun2022, Liao2023}. CLAP is analogous to likelihood-free inference approaches for parameter estimation and inference \citep[e.g.][]{Charnock2018, Fluri2021, Livet2021}, but instead of exploiting simulated data for inference, we perform KNN on real data and labels to ensure the probability density estimates to be locally calibrated. 

With CLAP as a reference, we delve into the miscalibration issue and showcase that it is a common shortcoming of conventional deep learning methods for probability density estimation. In particular, we present an illustration of miscalibration over different representative discriminative models, and reveal its association with uncertainties and correlations between data instances. In light of these findings, suggestions are given on circumventing miscalibration in deep learning in order to obtain reliable photometric redshift estimates applicable to actual astrophysical and cosmological analysis.

As a further note, another common issue for conventional methods is the mismatch between the spectroscopic sample used to train a model and the target (or test) sample to which the model is applied. As the learned distribution $p(z, d)$ is fixed in the model once the model is trained, applying it to a target sample with a different distribution would induce biases. In contrast, CLAP has the flexibility to match the target distribution by assigning different weights to the nearest neighbours for each data instance, offering a possible way to circumvent mismatches. This is another advantage of CLAP over conventional methods. We note that a mismatch may also lead to inconsistency between a model's confidence and accuracy for the target sample, though we use `miscalibration' to exclusively refer to the issue internal to the method implementation irrespective of mismatches. We will discuss the approaches built on CLAP for resolving mismatches in the next paper of our CLAP series, while in the current work we present the details of CLAP and solely focus on miscalibration.

In addition, we note that probability density estimates are denoted with probability density functions (PDFs) in most studies. From a rigorous statistical point, a PDF should characterise the dispersions intrinsic to a physical phenomenon, free from model assumptions, implementations, and noise in data. In actual applications, however, probability density estimates inevitably depend on the certain estimation method and the data quality. It is challenging to get rid of all those extra uncertainties, even after resolving miscalibration. We thus do not refer to probability density estimates as PDFs in order to avoid misuse of the terminology. For brevity, we adopt the expression $p(z|d)$ without adding extra terms to the condition unless otherwise noted.

This paper organised is as follows. Section~\ref{sec:data} describes the data used in this work, primarily galaxy images and spectroscopic redshifts. Section~\ref{sec:method} introduces our method CLAP. The information about the network architectures is provided in Appendix~\ref{sec:network}. In Sect.~\ref{sec:results} we show our results on probability density estimation using CLAP. More results and discussions are presented in Appendices~\ref{sec:examples}, \ref{sec:assess_recalibration}, \ref{sec:statistics_point}, and \ref{sec:spatial_nn}. An investigation on miscalibration is presented in Sect.~\ref{sec:miscalibration}. Finally, we summarise our results and give concluding remarks in Sect.~\ref{sec:conclusion}. In the section dedicated to Data and Code Availability, we provide the photometric redshift catalogues produced by CLAP and the code used in this work.

\section{Data} \label{sec:data}

\begin{figure*}
\sidecaption
\includegraphics[width=12cm]{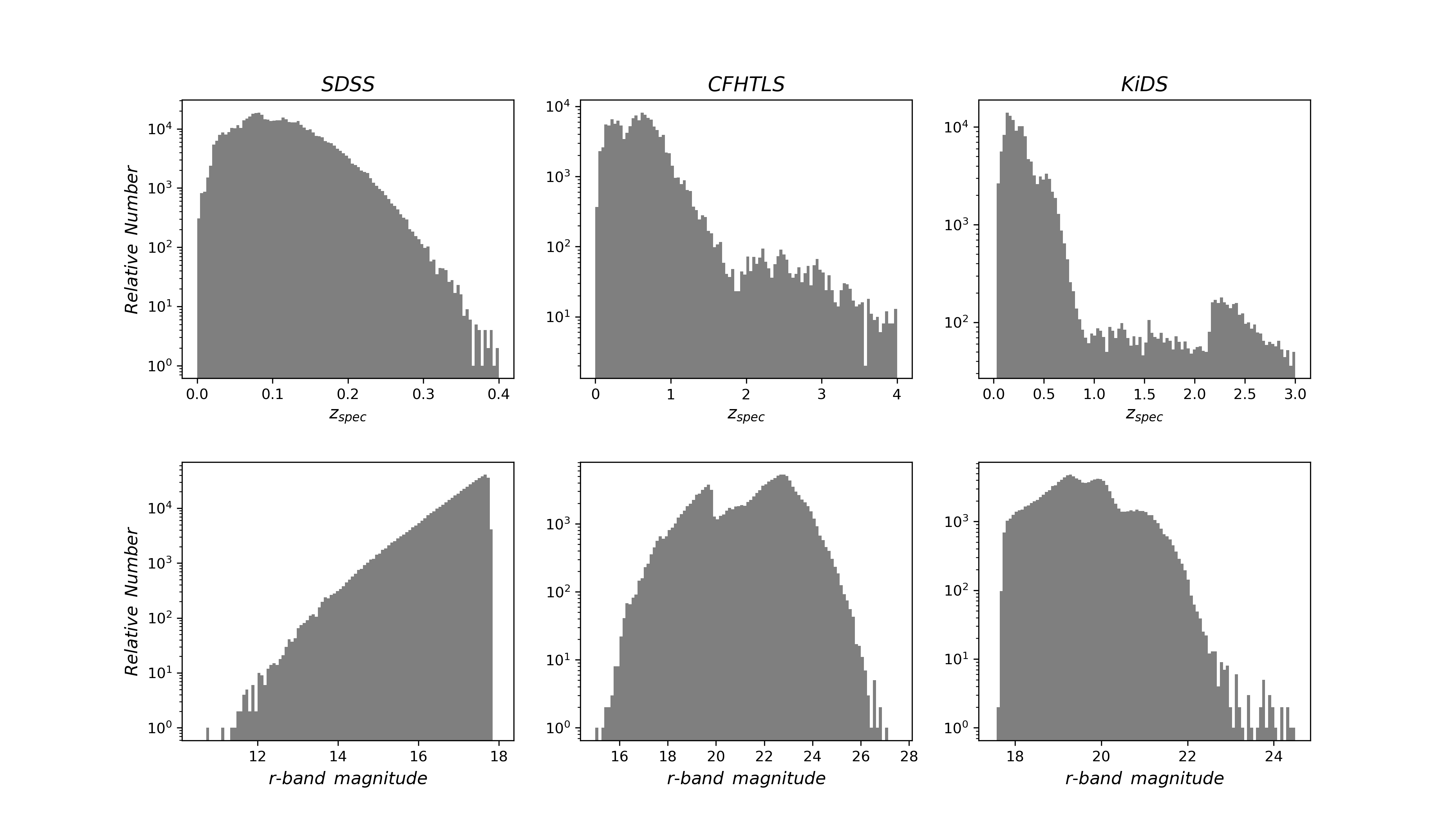}
\caption{Distributions of spectroscopic redshift and $r$-band magnitude for the SDSS, CFHTLS, and KiDS datasets used in our work.}
\label{fig:r_z_dist}
\end{figure*}

\begin{table*}\footnotesize
\renewcommand\arraystretch{1.5}
\caption{Data coverage in spectroscopic redshift and sample division.} \label{tab:coverage}
\centering
\begin{tabular}{c | c c c c c c c}
\hline
   &  $z_{min}$  &  $z_{max}$   &  Number of $z$ bins  &  Bin size  &  Training sample  &  Validation sample  &  Target sample  \\
\hline
SDSS  &  0.0  &  0.4  &  180  &  $2.22\times10^{-3}$  &  393\,219  &  20\,000  &  103\,305\tablefootmark{*}  \\
\hline
CFHTLS  &  0.0 &  4.0 &  1000  &  $4.0\times10^{-3}$  &  100\,000+23\,434\tablefootmark{**}  &  14\,759  &  20\,000  \\
\hline
KiDS  &  0.0 &  3.0 &  800  &  $3.75\times10^{-3}$  &  100\,000  &  14\,147  &  20\,000  \\
\hline
\end{tabular}
\tablefoot{
\tablefoottext{*}{50\% of the galaxies from the SDSS target sample are used as a reference subsample for the refitting procedure; the remaining 50\% are held out for inference (Sect.~\ref{sec:refitting}).}
\tablefoottext{**}{The 23\,434 galaxies with low-resolution spectroscopic redshifts from the CFHTLS training sample are used in supervised contrastive learning (SCL), but not in the downstream procedures.}
}
\end{table*}

We took data from three imaging surveys, the Sloan Digital Sky Survey (SDSS), the Canada-France-Hawaii Telescope Legacy Survey (CFHTLS), and the Kilo-Degree Survey (KiDS), each of which covers a different redshift and magnitude range. The distributions of spectroscopic redshift and $r$-band magnitude for each dataset are shown in Fig.~\ref{fig:r_z_dist}. The redshift coverage and sample division are detailed in Table~\ref{tab:coverage}.

\subsection{Sloan Digital Sky Survey (SDSS)}

The SDSS dataset used in this work contains 516\,524 galaxies from the Main Galaxy Sample available from SDSS Data Release 12 \citep{Alam2015}. These data were retrieved by \citet{Pasquet2019} and also used in \citet{Lin2022}, with detailed information provided in \citet{Pasquet2019}. This dataset has a low-redshift coverage $z<0.4$ and a cut at 17.8 on dereddened $r$-band petrosian magnitude, providing a rich reservoir of data in a restricted parameter space. The galactic reddening $E(B-V)$ for each galaxy along the line of sight is obtained using the dust map from \citet{Schlegel1998}. Each galaxy is associated with a spectroscopically measured redshift. Five stamp images that cover five optical bands $u,g,r,i,z$ were made to have the galaxy located at the centre and encompass $64\times64$ pixels in spatial dimensions, with each pixel covering 0.396 arcsec. The five images together with the galactic reddening $E(B-V)$ are regarded as a data instance and used as input data for CLAP. While previous work noticed that inputting the information of point spread functions (PSFs) to deep learning models would improve the estimation of galaxy properties \citep[e.g.][]{Umayahara2020, Li2022GaLNets}, we did not find a strong impact of PSFs on photometric redshift estimation for the data used in our work. Therefore, we did not include the PSF information in the input data for CLAP.

Via complete random sampling, we selected 393\,219 galaxies as a training sample, 20\,000 galaxies as a validation sample, and 103\,305 galaxies as a target sample, similar to the sample division by \citet{Pasquet2019}. Such division imposes that all the samples follow the same parent redshift-data distribution in spite of different sample sizes. This is the assumption we hold in this work without considering mismatches.

\subsection{Canada-France-Hawaii Telescope Legacy Survey (CFHTLS)}

We took the CFHTLS dataset that was described in detail in \Treyer{} and also used in \citet{Lin2022}. It contains 158\,193 galaxies observed in either the CFHTLS-Deep survey or the CFHTLS-Wide survey \citep{CFHTLST07}, with spectroscopic redshift up to $z \sim 4.0$ and dereddened $r$-band petrosian magnitude up to $r \sim 27.0$.

Since CFHTLS had no spectroscopic observations, the spectroscopic redshifts were retrieved from a few other spectroscopic surveys. The majority is a collection of high-quality redshifts measured with high S/N spectra or multiple spectral features, obtained from the COSMOS Lyman-Alpha Mapping And Mapping Observations survey \citep[CLAMATO; Data Release 1;][]{KGLee2018},
the DEEP2 Galaxy Redshift Survey \citep[Data Release 4;][]{Newman2013},
the Galaxy And Mass Assembly survey \citep[GAMA; Data Release 3;][]{Baldry2018},
the SDSS survey (Data Release 12),
the UKIDSS Ultra-Deep Survey \citep[UDS;][]{McLure2013, Bradshaw2013},
the VANDELS ESO public spectroscopic survey \citep[Data Release 4;][]{Garilli2021},
the VIMOS Public Extragalactic Redshift Survey \citep[VIPERS; Data Release 2;][]{Scodeggio2018},
the VIMOS Ultra-Deep Survey \citep[VUDS;][]{LeFevre2015},
the VIMOS VLT Deep Survey \citep[VVDS;][]{LeFevre2013},
the WiggleZ Dark Energy Survey \citep[Final Release;][]{Drinkwater2018}, and the zCOSMOS survey \citep{Lilly2007}. There is also a collection of low-resolution redshifts, acquired from the secure low-resolution prism redshift measurements from the PRIsm MUlti-object Survey \citep[PRIMUS; Data Release 1;][]{Coil2011, Cool2013} and the grism redshift measurements from the 3D-HST survey \citep[Data Release v4.1.5;][]{Skelton2014, Momcheva2016}. In the 158\,193 CFHTLS galaxies, 134\,759 have high-quality spectroscopic redshifts and 23\,434 have low-resolution redshifts. Similar to the SDSS images, the stamp images of each CFHTLS galaxy cover five optical bands $u,g,r,i,z$ and have $64\times64$ pixels with a pixel scale of 0.187 arcsec. The galactic reddening $E(B-V)$ is also included as a part of the data. 

We randomly selected 14\,759 galaxies as a validation sample and 20\,000 galaxies as a target sample, both from the high-quality collection. The remaining 100\,000 galaxies and the low-resolution collection of 23\,434 galaxies form a training sample. That is, for testing the results we only use the high-quality redshifts that are assumed to be secure, while the low-resolution redshifts are just used to increase statistics in training. In addition, we do not treat the CFHTLS-DEEP sample and the CFHTLS-WIDE sample separately as in \citet{Lin2022}.

\subsection{Kilo-Degree Survey (KiDS)}

We used the KiDS dataset retrieved by \citet{Li2022GaZNets} from the KiDS Data Release 4 \citep{Kuijken2019}. There are 134\,147 galaxies in this dataset, with spectroscopic redshift up to $z \sim 3.0$ and dereddened $r$-band `Gaussian Aperture and Point spread function (GAaP)' magnitude up to $r \sim 24.5$. The detailed description can be found in \citet{Li2022GaZNets}. Similar to the CFHTLS dataset, the spectroscopic redshifts of the KiDS galaxies were acquired from other surveys including the Chandra Deep Field South \citep[CDFS;][]{Szokoly2004}, the DEEP2 Galaxy Redshift Survey, the Galaxy And Mass Assembly survey (GAMA), and the zCOSMOS survey. The stamp images from KiDS cover four optical bands $u,g,r,i$ and have a pixel scale of 0.2 arcsec. Again, we set the spatial dimensions to be $64\times64$.

In addition to the KiDS images, this dataset includes the dereddened GAaP magnitudes in the five near-infrared (NIR) bands $Z,Y,J,H,Ks$ from the VISTA Kilo-degree Infrared Galaxy Survey \citep[VIKING;][]{Edge2014}, as well as the galactic reddening $E(B-V)$. Nonetheless, we refer to this dataset as `the KiDS dataset' for brevity throughout the paper.

Similar to \citet{Li2022GaZNets}, we randomly selected 100\,000 galaxies as a training sample, 14\,147 galaxies as a validation sample, and 20\,000 galaxies as a target sample.

\section{Our method: CLAP} \label{sec:method}

\begin{figure*}
\centering
\includegraphics[width=0.9\linewidth]{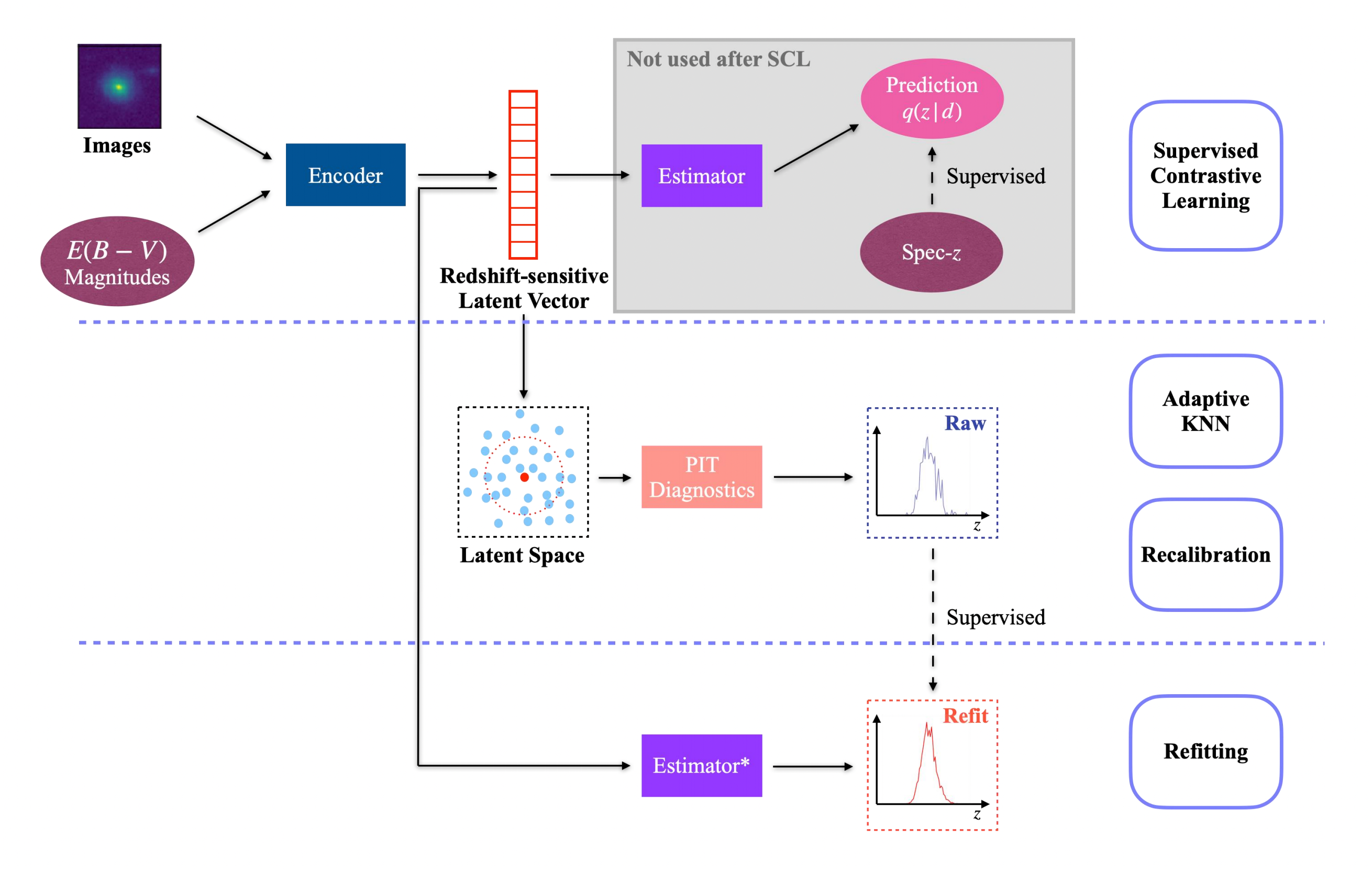}
\caption{Graphic illustration of our method CLAP for photometric redshift probability density estimation. The development of a CLAP model consists of several procedures, including supervised contrastive learning (SCL), adaptive KNN, reconstruction, and refitting. The SCL framework is based on neural networks. It uses an encoder network to project multi-band galaxy images and additional input data (i.e. galactic reddening $E(B-V)$, magnitudes) to low-dimensional latent vectors, which form a latent space that encodes redshift information and has a distance metric defined. The spectroscopic redshift labels are leveraged to supervise the redshift outputs predicted by an estimator network for extracting redshift information, but the trained estimator and its outputs are no longer used once the latent space is established (indicated by the shaded region). These outputs are uncalibrated, and should not be regarded as the final estimates produced by CLAP. The adaptive KNN and the KNN-enabled recalibration are implemented locally on the latent space via diagnostics with the probability integral transform (PIT), constructing raw probability density estimates using the known redshifts of the PIT-selected nearest neighbours. The raw estimates are then used as labels to retrain the estimator from scratch in a refitting procedure with the trained encoder fixed, resuming an end-to-end model ready to process imaging data and produce the desired estimates. Lastly, we combine the estimates from an ensemble of CLAP models developed following these procedures (not shown in the figure).}
\label{fig:clap}
\end{figure*}

\begin{figure*}
\centering
\includegraphics[width=\linewidth]{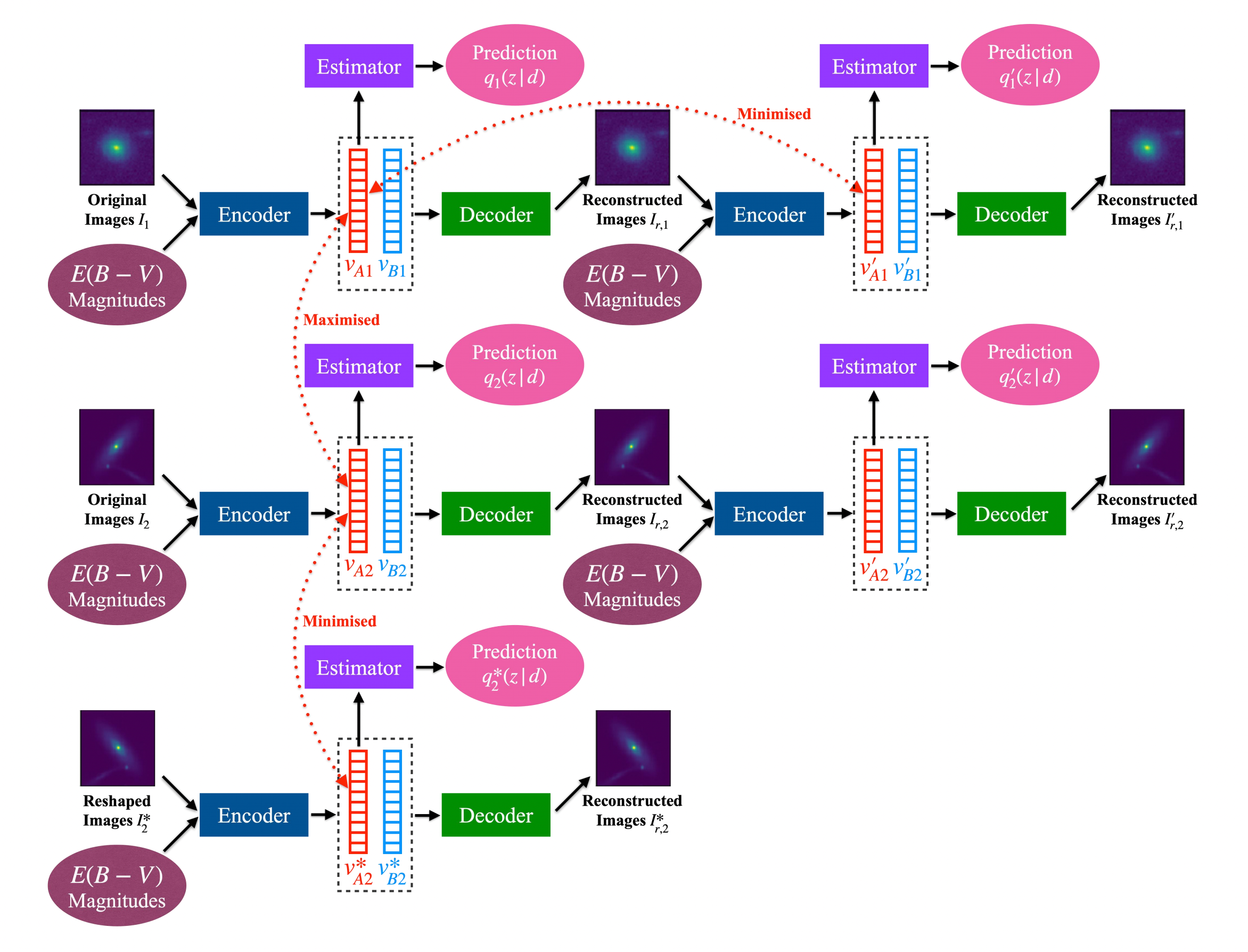}
\caption{Supervised contrastive learning (SCL) framework. It contains an encoder, an estimator, and a decoder. The encoder, the same as that shown in Fig.~\ref{fig:clap}, takes multi-band galaxy images and additional data as inputs, and produces two vectors $v_A$ and $v_B$. The vector $v_A$ is used to encode redshift information and is referred to as the `latent vector' throughout this work. It is inputted to the estimator that produces a redshift output supervised by the spectroscopic redshift label for extracting redshift information. The concatenation of $v_A$ and $v_B$ is inputted to the decoder to reconstruct images that resemble the input images. With the reconstructed images as inputs, this process is conducted again using the three networks with shared weights, producing the vector $v'_A$. Furthermore, the images reshaped with random flipping and rotation by 90 deg steps are exploited as inputs, producing the vector $v^*_A$. For contrastive learning, the contrast between $v'_A$ and $v_A$ and the contrast between $v^*_A$ and $v_A$ for the same galaxy are minimised (i.e. positive pairs), which are characterised by the Euclidean distance. The contrast between the latent vectors for any two different galaxies is maximised (i.e. a negative pair).}
\label{fig:scl}
\end{figure*}

\begin{algorithm}[H]
\textbf{Supervised contrastive learning:} Compress high-dimensional input data to low-dimensional redshift-sensitive latent vectors so that the subsequent KNN can be performed.

\textbf{Adaptive KNN:} Obtain the optimal number of nearest neighbours for each data instance via local PIT diagnostics, whose known spectroscopic redshifts are used to construct an initial probability density estimate.

\textbf{Recalibration:} Ensure that the obtained probability density estimates are locally calibrated.

\textbf{Refitting:} Resume an end-to-end model in order to bypass the expensive computation of KNN and regain the computational efficiency of deep learning.

\textbf{Combining an ensemble of estimates:} use the harmonic mean to combine the probability density estimates from an ensemble of models developed following the procedures above.
\caption{CLAP}
\label{alg:clap}
\end{algorithm}

\subsection{Overview} \label{sec:overview}

This section introduces our method CLAP. As illustrated in Fig.~\ref{fig:clap} and summarised in Algorithm~\ref{alg:clap}, a CLAP model is developed via supervised contrastive learning (SCL), adaptive KNN, reconstruction, and refitting, based on deep learning neural networks. For the model inputs, we prefer multi-band galaxy images rather than magnitudes or colours alone, because the images contain more information than photometry and potentially enable better accuracy to be achieved in photometric redshift estimation. On the other hand, as we demonstrate, the KNN algorithm is a key component of CLAP in resolving miscalibration, while the multi-band images have a large number of dimensions such that directly implementing the KNN on the images is infeasible. Therefore, CLAP first leverages contrastive learning to project the multi-band images and additional input data (i.e. galactic reddening $E(B-V)$, magnitudes) to low-dimensional redshift-sensitive latent vectors that form a latent space with a defined distance metric, enabling the subsequent KNN to be performed. It is essentially a deep learning-based compression of complex high-dimensional data. The contrastive learning is coupled with supervised learning using the spectroscopic redshift labels for better extraction of redshift information. 

The adaptive KNN is implemented on the obtained latent space. For each data instance from the target sample or the validation sample, the optimal $k$ value (i.e. the number of the nearest neighbours), which defines its neighbourhood, is determined via diagnostics based on the local PIT distributions. The PIT values are computed using the known spectroscopic redshifts of the selected nearest neighbours from the training sample. Once the $k$ value is determined, the known redshifts of the $k$ nearest neighbours within the neighbourhood are used to construct a probability density estimate, which is then recalibrated via local PIT diagnostics on the nearest neighbours again. The PIT values used for recalibration may be computed using the known redshifts from the validation sample. In essence, with the adaptive KNN and the KNN-enabled recalibration, the joint distribution $p(z, d)$ for the given dataset is modelled locally by leveraging the neighbourhood of the projection $\Phi(d)$ in the latent space, where $\Phi$ stands for the implemented model for projecting data in SCL that has been omitted in the expression $p(z|d)$ for brevity. 

The computational complexity of KNN precludes its use for processing large amounts of data envisioned by future imaging surveys. In order to avoid running computationally expensive KNN for every data instance from a potentially large target sample, a refitting procedure is implemented to resume an end-to-end model ready to directly produce the desired probability density estimates given input images, which regains the efficiency of deep learning.

Finally, for reducing uncertainties and improving accuracy, we develop an ensemble of CLAP models following the procedures above and combine the estimates from the ensemble. We propose that using the harmonic mean is a proper way for performing such combination.

The details of CLAP are elaborated in the following subsections.

\subsection{Supervised contrastive learning} \label{sec:scl}

We adopted a contrastive learning technique to establish a latent space that encodes redshift information (see \citealt{HuertasCompany2023} for a review of contrastive learning in astrophysics). The essence of contrastive learning is to have positive pairs and negative pairs simultaneously, minimising the contrast between positive pairs and maximising the contrast between negative pairs. A naive choice of generating positive pairs would be to modify the images on the pixel level such as colour jittering, resizing, smoothing, and adding noise, such that contrastive learning is performed in an unsupervised manner \citep[e.g.][]{Hayat2021, Wei2022}. However, our initial experiments showed that unsupervised contrastive learning with such pixel-level operations failed to produce a good representation for redshift, presumably because redshift is complex high-order information in a multi-dimensional spectroscopic and photometric parameter space, and may have exquisite dependence on pixel intensities \citep{Campagne2020}. Therefore, we propose to incorporate supervised learning in the contrastive learning technique using spectroscopic redshift labels, so that meaningful redshift information can propagate from input data to the latent space. We also consider implementing a deep learning-based reconstruction of redshift-informed images to provide positive counterparts for the original images to avoid performing pixel-level operations.

In addition, we do not perturb input data to account for the noise in data (or `aleatoric' uncertainties) like that applied in MLPQNA \citep{Brescia2014, Cavuoti2015, Cavuoti2017}, because, again, this may destruct redshift information. In fact, the errors in the input data from the training sample can be encoded in the latent space in the training process, and then encapsulated in the obtained probability density estimates by KNN for the target sample. Assuming that the errors in the training data are representative of those in the target data, these errors can be well accounted for without any particular treatment. While we assume that the spectroscopic redshifts used in this work are noiseless (except the low-resolution redshifts from the CFHTLS dataset), the possible errors in the spectroscopic redshifts from the training sample can also be encapsulated in the same way.

With these considerations, we developed a supervised contrastive learning (SCL) framework customised for the given problem (illustrated in Fig.~\ref{fig:scl}). It consists of an encoder network, an estimator network, and a decoder network. The encoder is fed with multi-band galaxy images (denoted with $I$) and additional input data, and outputs two low-dimensional vectors $v_A$ and $v_B$. The vector $v_A$ is used to encode redshift information, which we refer to as the `latent vector' that is taken to form a latent space, while the vector $v_B$ is used to encode other information extracted from the input data needed for reconstructing images. The vector $v_A$ is inputted to the estimator that produces a redshift output having the form of a probability density (denoted with $q(z|d)$), which is expressed in a series of bins given by the softmax function applied on the last fully connected layer and trained in a classification setting (i.e. considering each redshift bin as a class). The vectors $v_A$ and $v_B$ are concatenated and inputted to the decoder, reconstructing images that resemble the original inputs $I$. The reconstructed images (denoted with $I_r$) together with the original additional input data are re-inputted to the encoder, producing vectors $v'_A$ and $v'_B$, and subsequently a redshift output $q'(z|d)$ and images $I'_r$. In this second pass, the encoder, the estimator, and the decoder all use shared weights. Furthermore, we inputted the images $I^*$ that are original but reshaped with random flipping and rotation by 90 deg steps, yet the redshift is not modified under the assumption of spatial invariance. Correspondingly, $v^*_A$, $v^*_B$, $q^*(z|d)$, and $I^*_r$ were obtained.

In this framework, we regarded the images $I_r$ and $I$ (or equivalently, the latent vectors $v'_A$ and $v_A$) for the same galaxy as a positive pair on condition that redshift information has been passed to $v_A$, $v'_A$, and $I_r$. $I^*$ and $I$ (or $v^*_A$ and $v_A$) for the same galaxy were regarded as another positive pair, though this would not extract meaningful redshift information but only disentangle spatially variant information such as galaxy morphology. Lastly, the original images $I_1$ and $I_2$ for any two different galaxies (or the corresponding latent vectors $v_{A1}$ and $v_{A2}$) were regarded as a negative pair.

We adopted the Euclidean distance as a metric to characterise the contrast between two latent vectors $v_{A1}$ and $v_{A2}$,
\begin{equation}
D_{Euclidean} (v_{A1}, v_{A2}) = \sqrt{\frac{1}{L} \sum_{l=1}^L \left| v_{A1,l} - v_{A2,l} \right|^2},
\label{eq:dist_euclidean}
\end{equation}
where the index $l$ runs over the $L$ dimensions of the latent vectors. For a mini-batch of $n$ data instances from the training sample, we obtained the latent vectors $\{v_{A,i}\}$, $\{v'_{A,i}\}$, and $\{v^*_{A,i}\}$, where the index $i$ runs over all the data instances. Both $(v'_{A,i}, v_{A,i})$ and $(v^*_{A,i}, v_{A,i})$ for each data instance were used as positive pairs, having $2n$ pairs in total. The similarity function for all the positive pairs was defined as
\begin{equation}
Sim_p = \exp \left(-\frac{1}{2n} \sum_{i=1}^n \left( D_{Euclidean} (v'_{A,i}, v_{A,i}) + D_{Euclidean} (v^*_{A,i}, v_{A,i}) \right) \right).
\label{eq:sim_p}
\end{equation}
We generated $n/2$ negative pairs from $\{v_{A,i}\}$ each having two different galaxies (denoted with $(v_{A,i_p}, v_{A,\tilde{i_p}})$). The similarity function for all the negative pairs was defined as
\begin{equation}
Sim_n = \exp \left(-\frac{2}{n} \sum_{(i_p, \tilde{i_p}) \in \{i\}} D_{Euclidean} (v_{A,i_p}, v_{A,\tilde{i_p}}) \right).
\label{eq:sim_n}
\end{equation}
We then defined the contrastive loss function as
\begin{equation}
\mathcal{L}_{Contrastive} = -\log \frac{Sim_p}{Sim_p + Sim_n}.
\label{eq:contrastive}
\end{equation}

We converted the spectroscopic redshifts into one-hot labels using the same binning as the softmax output, assuming no spectroscopic errors beyond the bin size. The same one-hot label (denoted with $y$) was used to supervise both $q(z|d)$ and $q'(z|d)$ via the cross-entropy loss function
\begin{equation}
\mathcal{L}_{CE,y} = -\frac{1}{n} \sum_{i=1}^n \sum_{j=1}^m y_{ij} \left( \log q_{ij} + \log q'_{ij} \right),
\label{eq:CE1}
\end{equation}
where the index $i$ runs over the $n$ data instances in the mini-batch; the index $j$ runs over the $m$ redshift bins for each data instance. We also used the cross-entropy to impose mutual consistency between $q(z|d)$, $q'(z|d)$, and $q^*(z|d)$, 
\begin{equation}
\begin{aligned}
\mathcal{L}_{CE,q} = &-\frac{1}{n} \sum_{i=1}^n \sum_{j=1}^m \left( \hat{q}_{ij} \log q'_{ij} + \hat{q'}_{ij} \log q_{ij} \right)\\
&-\frac{1}{n} \sum_{i=1}^n \sum_{j=1}^m \left( \hat{q}_{ij} \log q^*_{ij} + \hat{q^*}_{ij} \log q_{ij} \right),
\end{aligned}
\label{eq:CE2}
\end{equation}
where the hat $\hat{}$ on top of $q(z|d)$, $q'(z|d)$, and $q^*(z|d)$ refers to gradient stopping in training when they are used as labels in front of the $\log$ terms. Equations~\ref{eq:CE1} and \ref{eq:CE2} ensure redshift information to be encoded in $v_A$, $v'_A$, $v^*_A$, and $I_r$.

We adopted the pixel-wise mean square error (MSE) to constrain the reconstructed images $I_r$, $I'_r$, and $I^*_r$,
\begin{equation}
\mathcal{L}_{MSE,I} = \frac{1}{n} \frac{1}{K} \sum_{i=1}^n \sum_{k=1}^K \left( \left| I_{r,k} - I_k \right|^2 + \left| I'_{r,k} - I_k \right|^2 + \left| I^*_{r,k} - I^*_k \right|^2 \right),
\label{eq:MSE}
\end{equation}
where the index $k$ runs over a total number of $K$ pixels of the images over multiple bands. The different MSE constraints for $I_r$ and $I^*_r$ ensure spatially variant information to be encoded in the vector $v_B$. We note that $I_r$ is required to be not identical to $I$, otherwise $v_A$ and $v'_A$ would collapse to a single point in the latent space, making the contrast between any different data instances uncontrollable and nullifying the contrastive learning. This requirement is naturally fulfilled since the random noise on $I$ cannot be fully recovered given a compression of information by the encoder.

Finally, we had the total loss function to perform SCL by summing up all the aforementioned loss functions,
\begin{equation}
\mathcal{L}_{SCL} = \mathcal{L}_{Contrastive} + \lambda_{CE} \left( \mathcal{L}_{CE,y} + \mathcal{L}_{CE,q} \right) + \lambda_{MSE,I} \mathcal{L}_{MSE,I},
\label{eq:loss_scl}
\end{equation}
where the coefficients $\lambda_{CE}$ and $\lambda_{MSE,I}$ define the relative weights among different terms. With a trained SCL framework, we can obtain the projection of each data instance (i.e. the latent vector) from a sample and establish a latent space.

As an additional note, having the estimator produce redshift outputs $q(z|d)$ in SCL is not yet to make estimation but only to let redshift information propagate to the latent vectors. These outputs are not used once the latent space is established. As we show, they are typically not calibrated, and should not be confused with the final estimates given by the complete implementation of CLAP that we introduce later, but they are compared in our analysis.

\subsection{Adaptive KNN} \label{sec:adaknn}

We performed KNN to retrieve a group of nearest neighbours for the projection of any query data instance in an established latent space, whose known redshifts were used to construct an initial probability density estimate. This idea is based on the topology of the latent space, that is, the data instances with neighbouring projections would have similar redshift-related properties and can be approximated as random draws from a common redshift probability density. The neighbourhood for a data instance in the latent space is represented by its nearest neighbours. We note that a neighbourhood would typically encapsulate different redshifts; at the same time, different neighbourhoods would overlap, having shared nearest neighbours that provide the same redshifts. In other words, the topology of the latent space naturally depicts the many-to-many mapping between photometric data and redshift.

Since the data are not uniformly distributed, the conventional KNN with a constant $k$ value as performed by \citet{Wei2022} may not define the neighbourhood properly for every query data instance. Therefore, we proposed an `adaptive' KNN algorithm with varying $k$ values for different data instances. To be specific, we relied on local PIT diagnostics to determine the optimal $k$ for each data instance. The PIT for a data instance is defined as
\begin{equation}
\mathrm{PIT} = \int_0^{z_{spec}} p(z|d) \mathrm{d}z,
\label{eq:pit}
\end{equation}
where $z_{spec}$ stands for its spectroscopic redshift. If a neighbourhood is properly defined that best describes the local properties and gives well-calibrated probability density estimates, the distribution of the PIT values computed for the data instances within the neighbourhood should be uniform between 0 and 1 \citep[see also][]{Dey2021, Zhao2021, Dey2022b}. Therefore, the optimal $k$ can be found when the PIT distribution constructed by the corresponding number of nearest neighbours is closest to uniformity (i.e. via PIT diagnostics). 

The adaptive KNN algorithm consists of two rounds of grid-search from a grid of $k$ values. The first round is performed to find the $k$ nearest neighbours for each query data instance with each $k$ selected from the grid. The query data instances and the nearest neighbours are all from the training sample. We adopt the Euclidean distance for KNN, the distance metric used in SCL. The known spectroscopic redshifts of the $k$ nearest neighbours provide a probability density; then with the known redshift of the query data instance, a PIT value for each given $k$ can be computed (denoted with ${\mathrm{PIT}_k}$). The second round of search, for each query data instance from the target sample or the validation sample, is performed to find its $k$ nearest neighbours from the training sample with each given $k$. Then with the ${\mathrm{PIT}_k}$ values of the $k$ nearest neighbours obtained from the first round, a distribution of ${\mathrm{PIT}_k}$ can be constructed that represents the local PIT distribution for the query data instance within a neighbourhood whose scale is characterised by the given $k$.

To quantify the discrepancy between uniformity and the ${\mathrm{PIT}_k}$ distribution for each $k$, we adopted the 1-Wasserstein distance \citep{Villani2009, Panaretos2019} defined as
\begin{equation}
D_{\mathrm{1\text{-}Wasserstein}} (\mathrm{PIT}_k) = \int_0^1 {\left| F_{\mathrm{PIT}_k} (P) - F_{uniform} (P) \right| \mathrm{d}P},
\label{eq:d_w_pit}
\end{equation}
where $F_{\mathrm{PIT}_k} (P)$ and $F_{uniform} (P)$ stand for the cumulative versions of the normalised ${\mathrm{PIT}_k}$ distribution and the uniform distribution, respectively. The 1-Wasserstein distance mainly captures the deviation between the broad shapes of the two distribution and is insensitive to random fluctuations in the ${\mathrm{PIT}_k}$ distribution. Using it as a metric, we searched for the optimal $k$ for each query data instance from the target sample or the validation sample, for which the corresponding $D_{\mathrm{1\text{-}Wasserstein}} (\mathrm{PIT}_k)$ reaches the minimum. Then the spectroscopic redshifts of the $k$ nearest neighbours were collected to construct an initial probability density estimate $p_{ini}(z|d)$, using the same binning as the softmax output in SCL.

\subsection{Recalibration} \label{sec:recalibration}

A PIT distribution closest to uniformity is not necessarily uniform. To ensure good calibration of the to-be-obtained probability density estimates, after determining the optimal $k$ and constructing $p_{ini}(z|d)$, we performed a (first-order) recalibration procedure similar to \citet{Bordoloi2010} using the local ${\mathrm{PIT}_k}$ distribution for the given $k$,
\begin{equation}
p_r(z|d) \propto p_{ini}(z|d) \times n_{\mathrm{PIT}_k} (F_{ini}(z|d)),
\label{eq:recalibration}
\end{equation}
where $p_r(z|d)$ is the recalibrated probability density estimate; $F_{ini}(z|d)$ stands for the cumulative density estimate given by $p_{ini}(z|d)$; $n_{\mathrm{PIT}_k} (F_{ini}(z|d))$ stands for the ${\mathrm{PIT}_k}$ number density at $F_{ini}(z|d)$. We note that this is a local recalibration procedure for every data instance enabled by KNN, rather than global calibration over the whole sample as performed by \citet{Bordoloi2010}.

A robust recovery of $n_{\mathrm{PIT}_k}$ is crucial for recalibration. Naturally, $n_{\mathrm{PIT}_k}$ can be approximated using the training sample alone,
\begin{equation}
n_{\mathrm{PIT}_k} \sim n_{\mathrm{PIT}_k,train}.
\label{eq:n_pit1}
\end{equation}
Given possible overfitting in SCL, the distributions of the projections $p(\Phi(d)|z)$ for the data seen and unseen in the training process may mildly differ, thus $n_{\mathrm{PIT}_k,train}$ may not perfectly reflect the actual PIT distribution within the defined neighbourhood of an unseen data instance. Therefore, we re-computed the $\mathrm{PIT}_k$ value for each data instance from the training sample, but with $z_{spec}$ in Eq.~\ref{eq:pit} being the redshift of its nearest instance from the validation sample. The distribution of such $\mathrm{PIT}_k$ values is denoted with $n_{\mathrm{PIT}_k,val}$. With this, we propose another estimate of $n_{\mathrm{PIT}_k}$ as
\begin{equation}
n_{\mathrm{PIT}_k} \sim \frac{1}{2} \left( n_{\mathrm{PIT}_k,train} + n_{\mathrm{PIT}_k,val} \right).
\label{eq:n_pit2}
\end{equation}
Combining $n_{\mathrm{PIT}_k,train}$ and $n_{\mathrm{PIT}_k,val}$ is due to the consideration that using $n_{\mathrm{PIT}_k,val}$ alone would introduce large errors because of limited statistics. For each dataset, we checked the recalibration results with either Eq.~\ref{eq:n_pit1} or Eq.~\ref{eq:n_pit2} using the validation sample, and applied the better one for the target sample. The KNN-calibrated raw probability density estimates $p_r(z|d)$ for the target sample produced by Eq.~\ref{eq:recalibration} are used for refitting described in the next subsection.
 
Essentially, this recalibration procedure determines the weights assigned to the nearest neighbours of each data instance that better characterise the local redshift-data distribution $p(z, d)$. An assessment on recalibration is presented in Appendix~\ref{sec:assess_recalibration}.

\subsection{Refitting} \label{sec:refitting}

To bypass the intensive computation required for KNN, we implemented a refitting procedure that resumed an end-to-end discriminative model ready to be applied directly to imaging data and produce probability density estimates $p(z|d)$. To be specific, for a SCL framework developed in Sect.~\ref{sec:scl}, we discarded the decoder, fixed the trained encoder and refit the estimator from scratch in a simple supervised manner with the KNN-calibrated raw probability density estimates $p_r(z|d)$ as labels. This is equivalent to retraining only the estimator using the obtained latent vectors as inputs. Both the inputs and the KNN-calibrated labels for refitting are now from the target sample. We note that the KNN-calibrated labels have no access to the actual spectroscopic redshifts from the target sample, thus there is no leakage of target redshift information to the model.

To better constrain the shapes of the probability density estimates $p(z|d)$, we adopted a hybrid loss function that consists of the cross-entropy and the 1-Wasserstein distance between $p(z|d)$ and $p_r(z|d)$,
\begin{equation}
\mathcal{L}_{Refit} = \frac{\lambda_{Refit}}{n} \sum_{i=1}^n \left( -\sum_{j=1}^m p_{r,ij} \log p_{ij} \sum_{j=1}^m \left| F_{ij} - F_{r,ij} \right| \Delta z \right),
\label{eq:loss_refit}
\end{equation}
where the index $i$ runs over the $n$ data instances in a mini-batch; the index $j$ runs over the $m$ redshift bins for each data instance. $F_{ij}$ and $F_{r,ij}$ stand for the two cumulative density estimates. $\Delta z$ is the size of a redshift bin (i.e. $\Delta z = (z_{max} - 0) / m$). The coefficient $\lambda_{Refit}$ controls the overall amplitude of the loss.

The model resumed this way has no difference from conventional end-to-end discriminative models but its outputs have been calibrated thanks to KNN. We demonstrate that refitting using the KNN-calibrated labels rather than the one-hot labels does not introduce apparent biases (Sect.~\ref{sec:assess_refitting}) or miscalibration (Sect.~\ref{sec:miscalibration}).

The output probability density estimates may be unfavourably discretised due to sparse data coverage in redshift bins. Hence, once a model was resumed, each output estimate was further smoothed with a Gaussian filter, whose dispersion was determined as 5\% of the standard deviation of the unsmoothed estimate. We found that this choice would effectively reduce the degree of discretisation but have negligible impact on the width of each probability density estimate (see Appendix~\ref{sec:examples} for a few examples).

In practice, when a target sample is too large to perform KNN, we can select a representative subset from the target sample as a reference subsample, implement the refitting procedure, and apply the resumed model to the remaining data for inference. We test the outcomes of inference using the SDSS dataset in Sect.~\ref{sec:assess_refitting}.

\subsection{Combining probability density estimates} \label{sec:combine}

Combining probability density estimates from an ensemble of measurements would reduce uncertainties and improve accuracy \citep{Treyer2024}. \citet{Pasquet2019} took the arithmetic mean over individual probability density estimates from six realisations of the same network to obtain the final estimates. Though seemingly intuitive, such combination tends to produce overdispersed probability density estimates. This is because $p(z|d,\{\Phi_i\}) \propto \frac{1}{n} \sum_{i=1}^n p(z|d,\Phi_i)$ does not hold true in general for any models $\{\Phi_i\}$ to be combined (see Sect.~\ref{sec:HMvsAM} for more discussions).

In this work, we suggest that a proper way to combine estimates from an ensemble of models is via the harmonic mean. To prove this, we first assume that the projection of a data instance $d$ (with a particular redshift $z$) via any individual model $\Phi_i$ to the latent space (i.e. $\Phi_i(d)$) is a random draw from the hypothetical neighbourhood of the projection via the combined model $\Phi_c$, though these individual projections do not need to be enclosed in an actual common neighbourhood. We also assume that the probability of getting the projection $\Phi_c(d)$ can be computed by averaging the probabilities of getting each individual projection $\Phi_i(d)$ under the linear approximation,
\begin{equation}
p(\Phi_c(d)) \propto \frac{1}{n} \sum_{i=1}^n p(\Phi_i(d)),
\label{eq:deriveHM1}
\end{equation}
or equivalently,
\begin{equation}
\frac{p(z,\Phi_c(d))}{p(z|\Phi_c(d))} \propto \frac{1}{n} \sum_{i=1}^n \frac{p(z,\Phi_i(d))}{p(z|\Phi_i(d))}.
\label{eq:deriveHM2}
\end{equation}
Since the data instance $d$ with the given redshift $z$ falls in the neighbourhoods of all the projections $\{\Phi_i(d)\}$ and $\Phi_c(d)$, the probability of $z$ belonging to any neighbourhood remains constant, i.e. $p(\Phi_c(d)|z) = p(\Phi_i(d)|z)$ for any model $\Phi_i$. This leads to the conclusion that
\begin{equation}
\frac{1}{p(z|d,\Phi_c)} \propto \frac{1}{n} \sum_{i=1}^n \frac{1}{p(z|d,\Phi_i)},
\label{eq:deriveHM3}
\end{equation}
for which $p(z|\Phi_c(d))$ and $p(z|\Phi_i(d))$ have been rewritten as $p(z|d,\Phi_c)$ and $p(z|d,\Phi_i)$, respectively, where $z$ is now considered as any redshift bins. Therefore, we propose to use the harmonic mean for combining an ensemble of probability density estimates. The combined estimates should be renormalised to unity if using the harmonic mean. For brevity in the text, we do not mention this explicitly hereafter. We show a few examples of probability densities before and after combination in Appendix~\ref{sec:examples}.

\subsection{Implementing CLAP} \label{sec:implement}

We implemented CLAP following all the procedures introduced in the previous subsections, using the SDSS, CFHTLS, and KiDS datasets separately. With each dataset, we used the training sample and the validation sample to develop the SCL framework, then included the target sample in the downstream procedures. For the CFHTLS dataset, both the galaxies with high-quality and low-resolution spectroscopic redshifts from the training sample were used in SCL, while only those with high-quality spectroscopic redshifts were used after SCL. This removes the mismatch between the initial training sample and the target sample.

For the encoder and the estimator in SCL, we adopted a modified version of the inception network developed by \citet{Treyer2024}. It falls in the category of convolutional neural networks (CNNs), featured by a set of multi-scale inception modules \citep{Szegedy2015} accompanied with convolutional layers and pooling layers. They produce feature maps that are then flattened and fed into a set of fully connected layers for predicting redshifts. Between the first two fully connected layers in this network, we inserted two new parallel layers to produce the vectors $v_A$ and $v_B$. The whole part of the network before the vectors is regarded as the encoder. The last fully connected layers, only linked to $v_A$, form the estimator. For it, we discarded the regression branch and only kept the classification branch (with a softmax function) from the original implementation by \citet{Treyer2024}. The redshift bin size we used in the softmax output and the one-hot spectroscopic redshift labels for each dataset is listed in Table~\ref{tab:coverage}. The concatenation of $v_A$ and $v_B$ is fed into the decoder that has convolutional layers and bilinear interpolation for upsampling and reconstructing images. The latent vector $v_A$ has 16 dimensions for encoding redshift information, and $v_B$ has 512 dimensions. These numbers were chosen to ensure computational efficiency and no major loss of information. For the SDSS dataset and the CFHTLS dataset, we included the galactic reddening $E(B-V)$ as an extra channel to the multi-band images inputted to the encoder. For the KiDS dataset, the 5-band NIR magnitudes from VIKING were further included as extra channels. More details about the network architectures can be found in Appendix~\ref{sec:network}.

The raw SDSS and CFHTLS images have AB magnitude zeropoints of 22.5 and 30, respectively. The KiDS images were redefined to have a zeropoint of 25. In order to lessen the significance of the galaxy peak flux, we then rescaled the images from all the datasets using the formula
\begin{equation}
I = \begin{cases}
-\log (-I_0 + 1.0), \quad I_0 < 0\\
\log (I_0 + 1.0), \quad I_0 > 0
\end{cases}
\label{eq:rescaling}
,\end{equation}
where $I$ and $I_0$ stand for the rescaled and the original pixel intensities, respectively.

In Eq.~\ref{eq:loss_scl}, we set the coefficient $\lambda_{MSE,I}=100$ for the SDSS and KiDS datasets while $\lambda_{MSE,I}=1$ for the CFHTLS dataset. $\lambda_{CE}=1$ was set for all the datasets. With each dataset, we trained our SCL framework from scratch using the mini-batch gradient descent implemented with the default Adam optimiser \citep{Adam}. In each training iteration, a mini-batch of 64 images-label pairs was randomly selected from the training sample. The images were randomly flipped and rotated with 90 deg steps before being used as the initial inputs $I$. We adopted an initial learning rate of $10^{-4}$, which was reduced by a factor of 5 after 60\,000 training iterations. By monitoring the loss with the validation sample, we decided to conduct 150\,000, 120\,000, 120\,000 training iterations in total for the SDSS, CFHTLS, and KiDS datasets, respectively. The SCL framework for each dataset was properly trained with these numbers, while we note that mild underfitting or overfitting would not have a significant impact on the results.

For the adaptive KNN and the recalibration procedures, we used 100 bins from 0 to 1 to express the PIT distributions for all the datasets. We note that a small number of bins would enlarge quantisation errors, while a large number of bins would make the data insufficient to fill in the bins, leading to shot noise. An intermediate number such as 100 is a compromise between reducing quantisation errors and limiting shot noise. For efficient grid search, we selected discrete $k$ values from four consecutive ranges: 5 to 200 with a step of 5, 200 to 600 with a step of 10, 600 to 1000 with a step of 20, and 1000 to 2000 with a step of 50. The maximum $k$ is set to be 2000, large enough to avoid finding too small (sub-optimal) $k$ values while still ensuring locality of each obtained neighbourhood and circumventing unnecessary intensive computation. The medians of the distributions of the obtained optimal $k$ are 660, 110, and 145 for the SDSS, CFHTLS, and KiDS target samples, respectively. The majority of data have optimal $k$ well below 2000, while only a minor fraction of the obtained optimal $k$ values (2.9\%, 0.2\%, and 0.5\%, respectively) are equal to this maximum value and thus may potentially exceed the defined $k$ range. We note that small $k$ values may be disfavoured because of large shot noise, but this potential bias can be corrected by recalibration (see Appendix~\ref{sec:assess_recalibration}).

In order to obtain the PIT distributions $n_{\mathrm{PIT}_k}$ for recalibration, by checking the validation samples, we chose to use Eq.~\ref{eq:n_pit2} for the SDSS and CFHTLS datasets, and Eq.~\ref{eq:n_pit1} for the KiDS dataset. We performed the second-order polynomial fit to remove random fluctuations in $n_{\mathrm{PIT}_k}$ and used the fit $n_{\mathrm{PIT}_k}$ in Eq.~\ref{eq:recalibration} for recalibration.

In the refitting procedure, in order to test whether the resumed end-to-end models are capable of making inference for held-out data, we split the SDSS target sample into a reference subsample and an inference subsample, used for refitting and inference, respectively, while the CFHTLS and KiDS target samples are only used as reference samples due to limited statistics. We set $\lambda_{Refit}=100$ in Eq.~\ref{eq:loss_refit}. For each dataset, we set the mini-batch size to be 128, took 40\,000 training iterations in total, and adopted a constant learning rate of $10^{-4}$. The mini-batch gradient descent with the default Adam optimiser was used again.

For the hyperparameters and coefficients involved in the implementation of CLAP, we tested a few different values around the aforementioned chosen values and found negligible impact on the results, indicating that the CLAP models established with the current implementation have converged. We refer to \citet{Lin2022} for detailed discussions on the impacts of varying implementing conditions, involving, for example, the training sample size, the redshift bin size, the number of training iterations, and the mini-batch size.

Finally, we leveraged the strategy of combining an ensemble of models to improve accuracy. Following the aforementioned procedures, we implemented CLAP ten times. Each time, the SCL and the refitting were conducted in the same way but with a different initialisation of trainable weights and different selections of mini-batches. The outcomes of adaptive KNN and recalibration may consequently be different due to such stochasticity. The output probability density estimates from the ten CLAP models were combined using the harmonic mean. 

We compared the models developed with the full range of data for each dataset illustrated in Fig.~\ref{fig:r_z_dist} and those developed with segmented ranges of redshift and $r$-band magnitude. The results produced by these `sub-models' generally follow the evolution trends of biases and accuracy obtained by the `full-models' shown in Sect.~\ref{sec:point_est}. Therefore, we only discuss the results from the full-models in the remainder of the paper.

\begin{table*}
\renewcommand\arraystretch{2.5}
\caption{Estimates and summary statistics derived from probability densities.} \label{tab:metrics}
\centering
\begin{tabular}{l | l}
\hline
Point estimate $z_{photo}$  &  $\int_0^{z_{max}} z \times p(z|d) \mathrm{d}z$ \\
\hline
Cumulative density estimate $F(z|d)$  &  $\int_0^{z} p(z'|d) \mathrm{d}z'$ \\
\hline
Instance-wise residual $\delta z$  &  $(z_{photo} - z_{spec})/(1 + z_{spec})$ \\
\hline
Bin-wise mean redshift residual $\delta_{<z>}$  &  $(<z_{photo}> - <z_{spec}>)/(1 + <z_{spec}>)$ \tablefootmark{*} \\
\hline
MAD-based deviation $\sigma_{\mathrm{MAD}}$ \tablefootmark{**}  &  $1.4826 \times \mathrm{Median} \left| \delta z - \mathrm{Median}(\delta z) \right| $ \\
\hline
Probability integral transform (PIT)  &  $\int_0^{z_{spec}} p(z|d) \mathrm{d}z$  \\
\hline
1-Wasserstein distance with the one-hot label &  $\int_0^{z_{spec}} F(z|d) \mathrm{d}z + \int_{z_{spec}}^{z_{max}} \left| 1 - F(z|d) \right| \mathrm{d}z$  \\
\hline
Continuous ranked probability score (CRPS)  &  $\int_0^{z_{spec}} F(z|d)^2 \mathrm{d}z + \int_{z_{spec}}^{z_{max}} \left( 1 - F(z|d) \right)^2 \mathrm{d}z$  \\
\hline
Cross-entropy with the one-hot label  &  $ -\log p(z_{spec}|d)$ \\
\hline
Entropy  &  $ -\int_0^{z_{max}} p(z|d) \log p(z|d) \mathrm{d}z$ \\
\hline
Standard deviation $\sigma$ &  $ \sqrt{\int_0^{z_{max}} (z - z_{photo})^2 p(z|d) \mathrm{d}z}$ \\
\hline
Skewness &  $ \int_0^{z_{max}} (z - z_{photo})^3 p(z|d) \mathrm{d}z / \sigma^3$ \\
\hline
Kurtosis &  $ \int_0^{z_{max}} (z - z_{photo})^4 p(z|d) \mathrm{d}z / \sigma^4$ \\
\hline
\end{tabular}
\tablefoot{
\tablefoottext{*}{$<>$ denotes the mean computed in a redshift or magnitude bin.}
\tablefoottext{**}{MAD stands for the median absolute deviation.}
}
\end{table*}

\begin{figure*}
\centering
\includegraphics[width=1.0\linewidth]{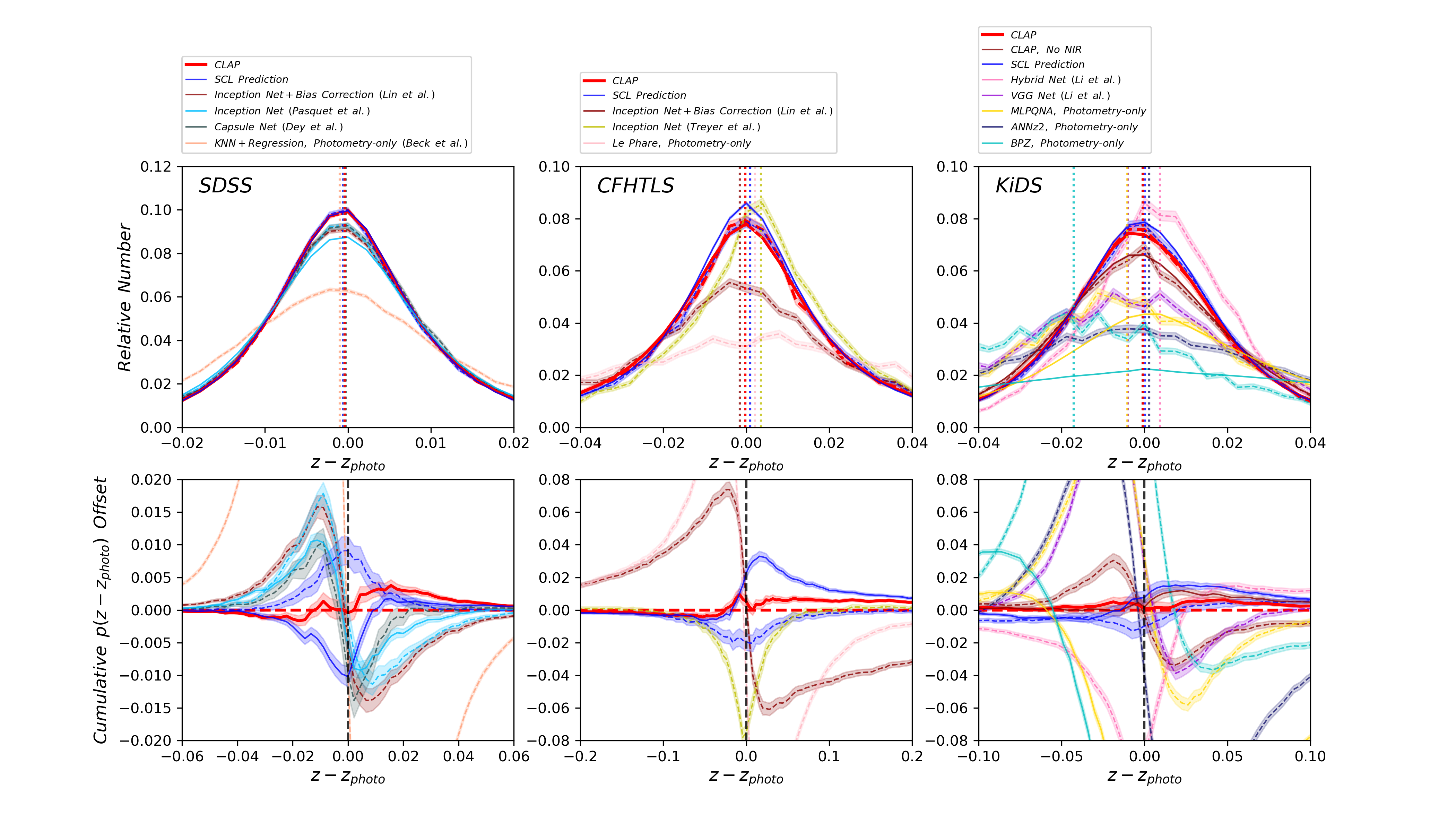}
\caption{Distribution offset analysis for the SDSS, CFHTLS, and KiDS target samples. The results obtained by CLAP are compared with the prediction from our supervised contrastive learning (SCL) framework, and those produced by a few benchmark image-based methods including inception networks from \citet{Pasquet2019} and \Treyer{}, network-based bias correlation from \citet{Lin2022}, a capsule network from \citet{Dey2022}, a VGG network, and a hybrid network from \citet{Li2022GaZNets}, and the photometry-only methods including KNN \& regression from \citet{Beck2016}, Le Phare \citep{Arnouts1999, Ilbert2006}, as well as MLPQNA \citep{Brescia2014, Cavuoti2015, Cavuoti2017}, ANNz2 \citep{Sadeh2016, Bilicki2018}, and BPZ \citep{BPZ2000} retrieved from the public KiDS catalogues. For the KiDS dataset, the results obtained by CLAP but without the NIR data are also shown (`No NIR'). For both CLAP and the SCL prediction, the ensemble of ten probability density estimates are combined using the harmonic mean. \textit{Upper panels:} $z - z_{photo}$ distributions. The solid curves show the stacked recentred probability densities $p_{\sum}(z - z_{photo})$ for the methods that have probability density estimation. The dashed curves show the $z_{spec} - z_{photo}$ histograms for all the methods. The error bands are estimated using bootstrap. The vertical dashed lines show the median of $z_{spec} - z_{photo}$ for each method, indicating the centre of each $z_{spec} - z_{photo}$ distribution regardless of the random errors due to limited statistics at high redshift. \textit{Lower panels:} Cumulative $z - z_{photo}$ distribution offsets. The solid curves show the cumulative offset of the stacked recentred probability density $p_{\sum}(z - z_{photo})$ relative to the corresponding $z_{spec} - z_{photo}$ histogram for each method that has a probability density estimation (i.e. $\Delta F_{p_{\sum}-h}$). The dashed curves show the cumulative offset of the $z_{spec} - z_{photo}$ histogram for each of all the methods (including No-NIR CLAP for the KiDS dataset) relative to that obtained by default CLAP (i.e. $\Delta F_{h-h}$). The error bands are estimated using bootstrap. The vertical dashed line in each panel indicates the zero point where $z$ and $z_{photo}$ coincide.}
\label{fig:main_pdfsc}
\end{figure*}

\begin{figure*}
\centering
\includegraphics[width=1.0\linewidth]{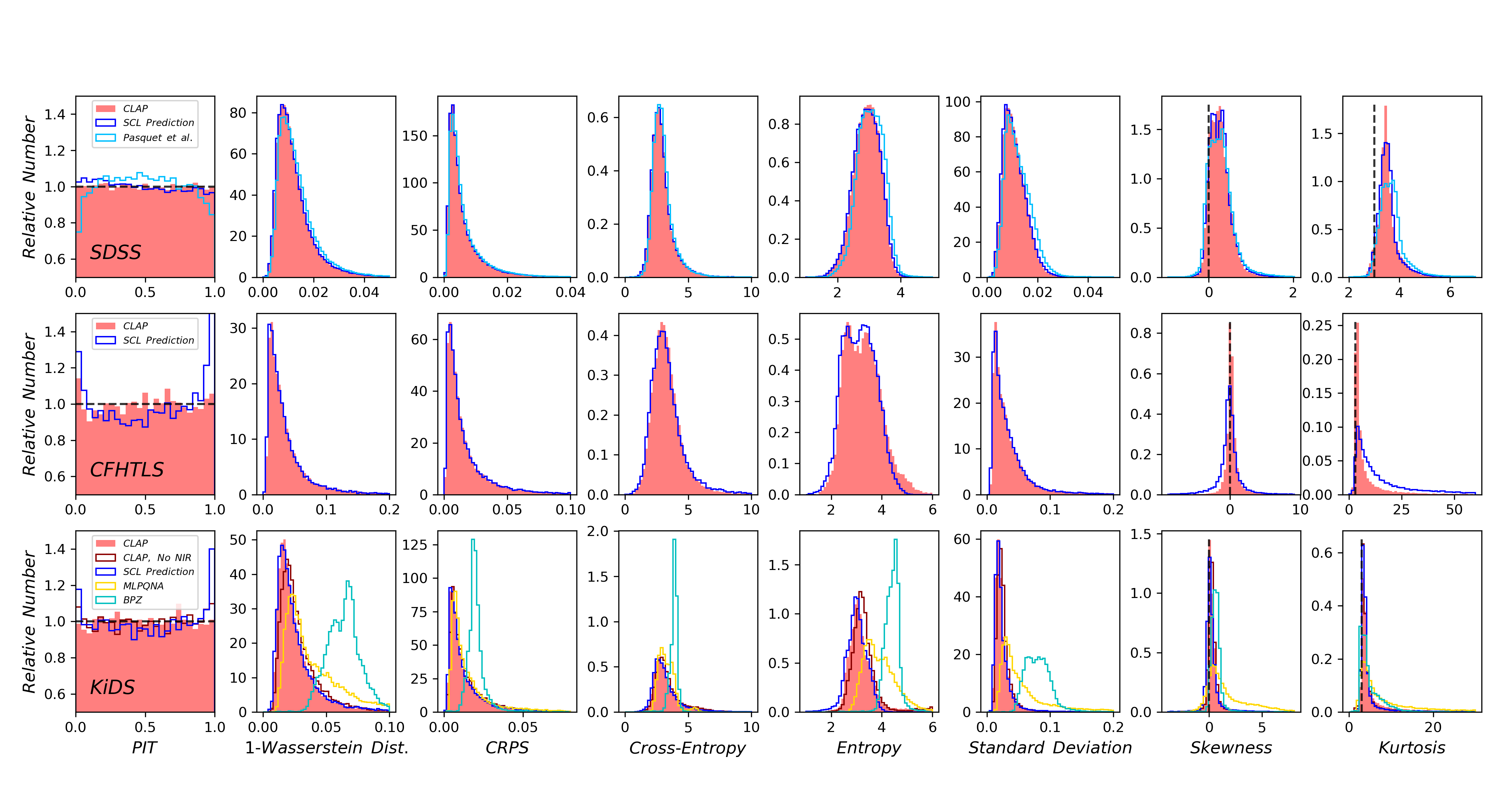}
\caption{Distributions of summary statistics of photometric redshift probability density estimates illustrated for the SDSS, CFHTLS, and KiDS target samples. Comparison is made among the results from CLAP and the methods for which the probability density estimates are available, as in Fig.~\ref{fig:main_pdfsc}. The definitions of the shown summary statistics can be found in Table~\ref{tab:metrics}. For both CLAP and the SCL prediction, the ensemble of ten probability density estimates are combined using the harmonic mean. For the PIT, the $y$-axis in each panel is truncated at 0.5 for clearer illustration; the horizontal dashed lines indicate normalised uniform distributions. The PIT distributions produced by MLPQNA and BPZ are dramatically non-flat and are not shown in the figure. For the kurtosis, the vertical dashed lines indicate the value for Gaussian-like distributions (i.e. 3).}
\label{fig:main_pdf_prop}
\end{figure*}

\section{Results on probability density estimation} \label{sec:results}

Throughout this section, we present our results on photometric redshift estimation by CLAP using the target samples from the SDSS, CFHTLS, and KiDS datasets. Comparisons are made with a few benchmark methods. In Sects.~\ref{sec:density_est}, \ref{sec:density_est2}, and \ref{sec:point_est}, we use the whole SDSS target sample without separating the reference and the inference subsamples, while in Sect.~\ref{sec:assess_refitting} we make this separation when assessing the resumption of end-to-end models by refitting. We demonstrate that the reference and the inference subsamples have consistent results produced by CLAP and thus they can be combined. A few examples of probability density estimates obtained by CLAP are presented in Appendix~\ref{sec:examples}.

\subsection{Assessment techniques} \label{sec:techniques}

Before further discussions in the next subsections, we present the techniques used for assessing the probability density estimates and the density-derived point estimates.

Firstly, we discuss our choice of point estimates, which is important for our analysis. There are several possible definitions of point estimates that can be derived from probability density estimates, such as $z_{mean}$ (i.e. the probability-weighted mean redshift), $z_{mode}$ (i.e. the redshift at the peak probability), and $z_{median}$ (i.e. the median redshift at which the cumulative probability is 0.5). Different definitions have been discussed in many studies such as \citet{Tanaka2018}, \citet{Pasquet2019}, \citet{Lin2022}, \citet{Treyer2024}, and \Treyer{}. For template-fitting methods, the best-fitting redshifts are usually adopted as point estimates, analogous to $z_{mode}$. In our work, we chose $z_{mean}$ as point estimates. Our consideration for this links back to the frequentist definition: if a probability density accurately describes the actual rate of occurrence of true redshifts, the empirical mean of the true redshifts should be equal to the expected mean redshift weighted by the probability density, i.e. $z_{mean}$. In other words, with sufficient statistics, perfect probability density estimates lead to zero mean redshift residuals if estimated with $z_{mean}$ (though the reverse is not necessarily true). Other choices of point estimates do not have such property in general. Therefore, $z_{mean}$ is taken as our exclusive choice of point estimates $z_{photo}$ for CLAP hereafter.

To have an inspection over the properties of the probability density estimates, we checked a few summary statistics defined using the individual probability densities and the corresponding spectroscopic redshifts $z_{spec}$, including the probability integral transform (PIT), the 1-Wasserstein distance, the continuous ranked probability score (CRPS), the cross-entropy, the entropy, the standard deviation, the skewness, and the kurtosis. Regarding the point estimates, we characterised the mean redshift biases using $\delta_{<z>}$, the residuals of mean redshifts estimated in redshift or magnitude bins, and quantified the estimation accuracy using $\sigma_{\mathrm{MAD}}$, the estimate of deviation based on the median absolute deviation (MAD). The definitions of all these metrics are shown in Table~\ref{tab:metrics}. 

Knowing that each individual probability density has only one realisation (i.e. a unique spectroscopic redshift for a given galaxy), we also considered a distribution offset analysis with a stacking technique to assess the global behaviour of the probability density estimates over a sample. Specifically, we recentred each probability density estimate by shifting the point estimate $z_{photo}$ to $z=0$ and stacked all such recentred estimates over a sample (denoted with $p_{\sum}(z - z_{photo})$); we also drew the histogram of $z_{spec} - z_{photo}$. If the probability density estimates are well-calibrated, these two distributions should be consistent given sufficient statistics (though the reverse is not necessarily true). Hence, we checked the cumulative offset of the stacked recentred probability density $p_{\sum}(z - z_{photo})$ relative to the $z_{spec} - z_{photo}$ histogram (denote with $\Delta F_{p_{\sum}-h}$). Such cumulative offset is sensitive to any inconsistency between the two distributions to be compared.

Although many studies stacked the original probability density estimates and made a comparison with the actual $z_{spec}$ distribution, we note that such stacked probability density and the $z_{spec}$ distribution are usually not directly comparable. In other words, for probability densities $p(z|d_1), p(z|d_2), ..., p(z|d_n)$ to be stacked, there may exist a data instance with any particular redshift $z$ enclosed in the neighbourhoods of $d_1, d_2, ..., d_n$ simultaneously, making the selections of $(z, d_1), (z, d_2), ..., (z, d_n)$ not mutually exclusive, such that $p(z, \cup_{i=1}^n d_i) \neq \sum_{i=1}^n p(z, d_i)$ \citep[see also][]{Malz_Alex2021}. However, the connections among $d_1, d_2, ..., d_n$ through the shared data instance with $z$ can be broken by instead using $\Delta z_i = z - z_{photo,i}$ for each $d_i$. This is because $(z, d_1), (z, d_2), ..., (z, d_n)$ that initially had the shared $z$ are now re-assigned to different $\{\Delta z_i\}$ bins since $\{z_{photo,i}\}$ are different. Now the selections of $(\Delta z, d_1), (\Delta z, d_2), ..., (\Delta z, d_n)$ for the same $\Delta z$ are statistically mutually exclusive, since $\Delta z$ is no longer contributed by a shared data instance in the neighbourhoods of $d_1, d_2, ..., d_n$. Hence, the stacked recentred probability density $p_{\sum}(z - z_{photo})$ and the $z_{spec} - z_{photo}$ distribution are now comparable provided with sufficient statistics. Admittedly, this stacking technique is not rigorously mathematically meaningful, but it serves as a tool to check the reliability of the probability density estimates globally over a sample.

Similarly, we checked the cumulative offset between the $z_{spec} - z_{photo}$ histograms for two methods (denote with $\Delta F_{h-h}$), which can be used to demonstrate their relative accuracy and biases, irrespective of whether or not they have probability density estimation.

\subsection{Benchmark methods} \label{sec:benchmark}

We compared CLAP with a few benchmark methods for the three datasets. The results predicted by our SCL framework (i.e. the softmax outputs $q(z|d)$ from SCL rather than refitting) are shown for the first comparison. We reiterate that the SCL prediction is only regarded as a baseline and should not be confused with the results from the complete implementation of CLAP. For the KiDS dataset, we also show the results produced by CLAP but without including the 5-band NIR magnitudes from VIKING as inputs, demonstrating the impact of the NIR data on photometric redshift estimation. For both CLAP and the SCL prediction, the ensemble of ten probability density estimates are combined using the harmonic mean unless otherwise noted. 

For further comparison, we present the results obtained by several image-based deep learning methods collected from different studies, including inception networks from \citet{Pasquet2019} and \Treyer{}, a bias correlation approach based on inception networks from \citet{Lin2022}, a capsule network from \citet{Dey2022}, as well as a VGG network (using the 4-band KiDS images as inputs) and a hybrid network (using the 4-band KiDS images and the 5-band VIKING photometry as inputs) from \citet{Li2022GaZNets}. Regarding photometry-only methods, we show the results from \citet{Beck2016} who applied KNN and local regression, the results obtained using Le Phare \citep{Arnouts1999, Ilbert2006}, as well as those retrieved from the public KiDS catalogues\footnote{\url{https://kids.strw.leidenuniv.nl/DR3/index.php}} produced by MLPQNA \citep{Brescia2014, Cavuoti2015, Cavuoti2017}, ANNz2 \citep{Sadeh2016, Bilicki2018}, and BPZ \citep{BPZ2000}.

We show the results from \citet{Lin2022}, \citet{Pasquet2019}, \citet{Dey2022}, \Treyer{}, and \citet{Li2022GaZNets} using their test samples. All but the CFHTLS test sample from \citet{Lin2022} essentially have the same redshift-data distributions as our target samples in spite of different sample divisions. We use the results provided by \citet{Lin2022} for which the $z_{photo}$-dependent mean redshift biases have been corrected, bypassing the impact of the mismatch in distributions. The results from \citet{Beck2016} and Le Phare are shown using our target samples, while those from MLPQNA, ANNz2, and BPZ are shown using the test sample provided by \citet{Li2022GaZNets}. These results are directly compared despite that different choices of point estimates were adopted by different studies. \citet{Pasquet2019}, MLPQNA, and BPZ have probability density estimates available for comparison, while the other methods only provide point estimates. The probability density estimates provided by \citet{Pasquet2019} have the same redshift coverage and binning as ours. For consistency with our analysis on the KiDS dataset selected with $z_{spec} < 3$, the original probability density estimates given by MLPQNA and BPZ were truncated at $z=3$, rebinned, and renormalised. We further caution that the VGG network, MLPQNA, ANNz2, and BPZ for the KiDS dataset did not use the five NIR bands, thus they can be treated as a direct reference for the `No-NIR' CLAP implementation, but not the default implementation with the NIR data included. These results are all shown to enrich the comparison among various methods.

\subsection{Calibration of probability density estimates} \label{sec:density_est}

In the upper panels of Fig.~\ref{fig:main_pdfsc}, the dashed curves show the $z_{spec} - z_{photo}$ histograms for CLAP and the benchmark methods for the SDSS, CFHTLS, and KiDS datasets. The solid curves show the stacked recentred probability densities $p_{\sum}(z - z_{photo})$ when the probability density estimates are available. Their cumulative offsets $\Delta F_{p_{\sum}-h}$ are shown as the solid curves in the lower panels. The first column of Fig.~\ref{fig:main_pdf_prop} shows the PIT distributions. The cumulative offset $\Delta F_{p_{\sum}-h}$ and the PIT distribution are complementary and offer a cross-check over the calibration of a given set of probability density estimates.
 
For all the three datasets, the cumulative offsets $\Delta F_{p_{\sum}-h}$ given by CLAP (including the No-NIR implementation for the KiDS dataset) are constrained within low levels (i.e. 0.005, 0.01, and 0.01, respectively), better than those given by the SCL prediction and the other benchmark methods. CLAP also produces nearly flat PIT distributions, though the PIT distributions obtained for the CFHTLS dataset and by the No-NIR implementation for the KiDS dataset are slightly convex, in addition to some fluctuations probably as a result of estimation errors. The exclusion of the NIR data for the KiDS dataset degrades the estimation accuracy (see Sect.~\ref{sec:point_est}), but hardly influences the calibration of probability densities. In particular, No-NIR CLAP outperforms the SCL prediction for which the NIR data were used. Both the small offsets and the nearly flat PIT distributions gain evidence that the probability density estimates obtained by CLAP are well-calibrated.

In contrast, for either the CFHTLS dataset or the KiDS dataset, the tilde around zero in the cumulative offset $\Delta F_{p_{\sum}-h}$ given by the SCL prediction indicates that the stacking of the recentred probability density estimates is narrower than the $z_{spec} - z_{photo}$ histogram, reflecting that the probability density estimates are underdispersed (or overconfident), as also implied by the convex PIT distributions. This is a sign of miscalibration caused internally by the estimation method. The probability density estimates given by MLPQNA and BPZ are significantly uncalibrated, producing large cumulative offsets $\Delta F_{p_{\sum}-h}$. For the SDSS dataset, the trough centred at zero in the cumulative offset $\Delta F_{p_{\sum}-h}$ given by the SCL prediction indicates that the probability density estimates are on average shifted towards higher redshift, resulting in overestimated $z_{photo}$ and the slightly tilted PIT distribution, though the PIT distribution is already close to flatness thanks to high accuracy. For \citet{Pasquet2019}, the reversed tilde and the concave PIT distribution indicate that the probability density estimates are overdispersed, which is an artefact erroneously introduced by combining probability density estimates using the arithmetic mean (discussed in Sect.~\ref{sec:HMvsAM}).

These results indicate that CLAP is more robust than the other methods in obtaining well-calibrated probability density estimates. We present a deeper investigation on miscalibration in Sect.~\ref{sec:miscalibration}. We note that the adaptive KNN algorithm and recalibration are both key parts of CLAP for ensuring good calibration for each data instance. An assessment on recalibration is presented in Appendix~\ref{sec:assess_recalibration}.

\subsection{Properties of probability density estimates} \label{sec:density_est2}

Figure~\ref{fig:main_pdf_prop} illustrates the distributions of the summary statistics other than PIT for the methods that have probability density estimation. For CLAP, the SCL prediction, and \citet{Pasquet2019}, the distributions of the 1-Wasserstein distance, the CRPS, the cross-entropy, and the standard deviation all have heavy tails towards large values, resulting from the wide probability densities and the large uncertainties at high magnitude. The distributions of these summary statistics obtained by No-NIR CLAP for the KiDS dataset are all slightly shifted towards higher values due to enlarged estimation errors compared to the results from default CLAP. Both the distributions of the cross-entropy and the entropy are peaked around 3, meaning that the probability at the redshift bin that $z_{spec}$ falls in is most likely to be $\exp(-3) \sim 0.05$, which is consistent with the probability that has the largest contribution to the entropy. The subtle double peaks in the entropy distribution for CFHTLS dataset reflects the two populations clearly seen in the $r$-band magnitude distribution in Fig.~\ref{fig:r_z_dist}, an artefact introduced by combining disparate spectroscopic surveys. The slightly excessive tails in the entropy distributions obtained by CLAP for the CFHTLS and KiDS datasets are due to the smoothing applied on the probability densities in the refitting procedure. In contrast to the image-based methods, MLPQNA and BPZ tend to produce higher values for these summary statistics due to lower accuracy and wider probability densities.

The bulk of the skewness distribution for the SDSS dataset is located at positive values, showing that the majority of probability density estimates are right skewed as a result of the skewed redshift distribution for the training sample. The skewness distributions for the CFHTLS and KiDS datasets are also centred above zero, whereas there are much heavier tails than those for the SDSS dataset. The tails of the kurtosis distributions are all above 3, indicating that the majority of probability density estimates have fatter tails than Gaussian-like distributions. Significantly higher values are present for the CFHTLS and KiDS datasets, reaching up to 30.9 and 56.4 at the 90th percentile, respectively. The variety of the skewness and the kurtosis reveals the diversity of photometric redshift probability densities estimated over broad redshift and magnitude ranges.

\begin{figure*}
\centering
\includegraphics[width=0.95\linewidth]{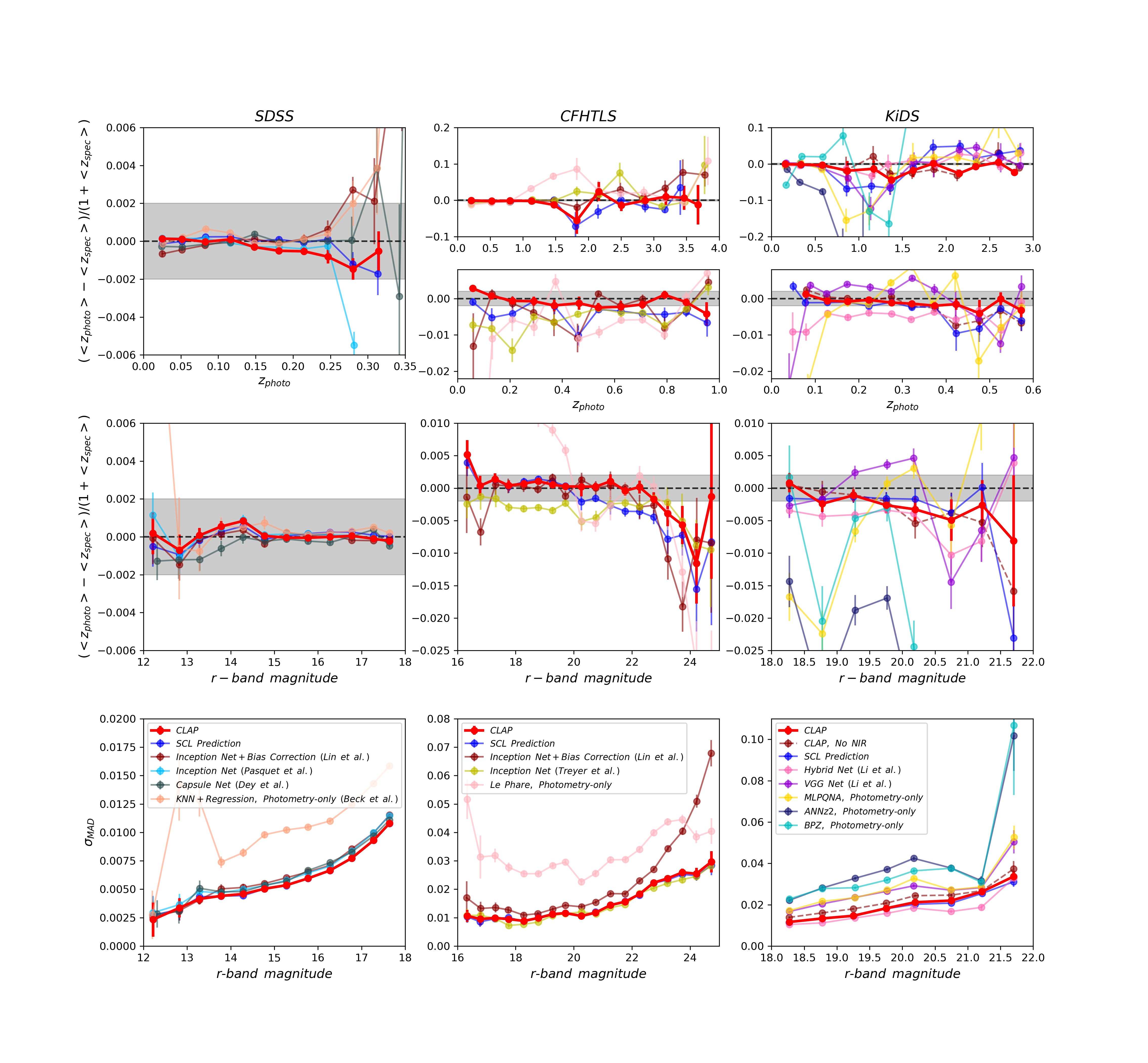}
\caption{Mean redshift residuals and accuracy of photometric redshift point estimates shown for the SDSS, CFHTLS, and KiDS target samples. The results obtained by CLAP are compared with those produced by other methods, as in Fig.~\ref{fig:main_pdfsc}. For both CLAP and the SCL prediction, the ensemble of ten probability density estimates are combined using the harmonic mean. \textit{Upper panels:} Mean redshift residuals as a function of photometric redshift $z_{photo}$. The error bars are estimated using the root mean square error (RMSE) of residuals in each photometric redshift bin. The requirement for the accuracy of photometric redshift estimation (i.e. $|\delta_{<z>}| < 0.002$) is indicated by the shaded bands. The horizontal dashed lines indicate zero residuals. The results for the CFHTLS and KiDS datasets are shown in both the full and the restricted redshift ranges; the restricted range (i.e. $z_{photo}<1.0$ and $z_{photo}<0.6$, respectively) roughly covers 90\% of the full sample for each dataset. For the KiDS dataset, the results produced by ANNz2 and BPZ are not shown in the restricted range due to large biases. \textit{Middle panels:} Same as the upper panels, but showing mean redshift residuals as a function of $r$-band magnitude for the full data. \textit{Lower panels:} $\sigma_{\mathrm{MAD}}$ as a function of $r$-band magnitude. The error bars are estimated using bootstrap.}
\label{fig:main_point_est}
\end{figure*}

\subsection{Mean redshift biases and accuracy of point estimates} \label{sec:point_est}

Using the point estimates derived from the probability density estimates, we evaluated the mean redshift biases and the accuracy, characterised by $\delta_{<z>}$ and $\sigma_{\mathrm{MAD}}$, respectively, which are important indicators of the quality of photometric redshift estimation. Accurate mean redshift determination is crucial for weak lensing tomography analysis \citep[e.g.][]{Ma2006, Huterer2006, Ilbert2021}. For modern cosmological surveys such as \textit{Euclid} and LSST, the error of the mean redshift determined in a tomographic bin is required to be no more than 0.002 \citep{Laureijs2011}. This is a demanding requirement that necessitates careful evaluation on the photometric redshift estimation results.

Figure~\ref{fig:main_point_est} illustrates how $\delta_{<z>}$ and $\sigma_{\mathrm{MAD}}$ evolve as a function of $z_{photo}$ or $r$-band magnitude for CLAP and the other methods. These results are also summarised in Appendix~\ref{sec:statistics_point}. As a further comparison, the dashed curves in the lower panels of Fig.~\ref{fig:main_pdfsc} show the cumulative offset of the $z_{spec} - z_{photo}$ histogram for each of all the methods (including No-NIR CLAP for the KiDS dataset) relative to that for default CLAP (i.e. $\Delta F_{h-h}$).

At first glance, the results obtained by the image-based methods are in general better than those obtained by the photometry-only methods. Most image-based methods achieve higher accuracy characterised by $\sigma_{\mathrm{MAD}}$, which might primarily result from the exploitation of information in addition to galaxy photometry. The lower accuracy obtained by No-NIR CLAP and the VGG network compared to that by the other image-based methods for the KiDS dataset indicates the need for the 5-band NIR magnitudes as inputs. We notice that there seems to be no significant difference among most image-based methods in constraining the mean redshift biases. In particular, the results from all the image-based methods shown for the SDSS dataset fulfil the requirement $|\delta_{<z>}| < 0.002$ over $z_{photo}<0.25$, associated with high accuracy due to the good quality of the magnitude-limited low-redshift SDSS data and a large training sample size. For the CFHTLS and KiDS datasets, although there are large scatters at high redshift and high magnitude due to limited statistics and low S/N, the mean redshift residuals produced by most image-based methods generally exhibit no drastic evolving trend as a function of $z_{photo}$ over the full redshift range. The photometry-only methods, on the contrary, typically produce higher $\sigma_{\mathrm{MAD}}$, as also implied by the strong tildes in their cumulative offsets $\Delta F_{h-h}$ relative to CLAP, though the accuracy achieved by MLPQNA is only mildly different from that by the VGG network. ANNz2 and BPZ produce much stronger biases. The mutually enhanced errors and biases may primarily stem from the photometry-redshift degeneracies and the photometric errors. The template-fitting methods (i.e. Le Phare and BPZ) may additionally suffer from a shortage of representative templates.

In spite of similar results given by the image-based methods, we point out that CLAP performs better in constraining the mean redshift biases given sufficient statistics. For each of the CFHTLS and KiDS datasets, we show the mean redshift residuals in a restricted redshift range (i.e. $z_{photo}<1.0$ and $z_{photo}<0.6$, respectively) in addition to the full range, which contains roughly 90\% of the full sample. As can be seen, the residuals produced by default CLAP are marginally consistent with the requirement $|\delta_{<z>}| < 0.002$, better than those by the conventional discriminative models such as the inception network from \Treyer{} and the hybrid network from \citet{Li2022GaZNets}. The outputs by these conventional methods at low redshift would probably be compromised in order to accommodate the large errors at high redshift. The non-zero residuals at low redshift might also be compounded by stochasticity in the training process and exhibit variations if there were multiple realisations of the same network (further discussed in Sect.~\ref{sec:mis_illustration}). Similar behaviours can be noticed for residuals shown as a function of $r$-band magnitude. This is in line with the fact that the $z_{spec} - z_{photo}$ histograms obtained by these methods all have centres slightly deviating from zero, as indicated by the vertical dotted lines overplotted on the $z_{spec} - z_{photo}$ histograms and also the troughs in the cumulative offsets $\Delta F_{h-h}$. The results given by the SCL prediction have similar but weaker behaviours. In contrast, No-NIR CLAP, though having lower accuracy, still produces a $z_{spec} - z_{photo}$ histogram centred at zero and has a good constraint on the mean redshift biases. For the SDSS dataset, we see a clear drop at around $z_{photo} \sim 0.25$ for the results from \citet{Pasquet2019}, which was also illustrated in Fig.~8 therein. The cumulative offset $\Delta F_{h-h}$ for the SCL prediction has a bump around zero, echoing with the trough in $\Delta F_{p_{\sum}-h}$, indicating that the $z_{photo}$ estimates given by the SCL prediction are on average shifted towards higher values.

In general, we note that the conventional discriminative models cannot always guarantee near-zero mean redshift residuals. This may be due to, for example, the joint impact of overfitting, data imbalance, and stochasticity in the training process, which are all related to the phenomenon of miscalibration. While the effect of miscalibration on the point estimates is insignificant for the SDSS dataset as indicated by the good bias constraints over $z_{photo}<0.25$, this is no longer the case for the CFHTLS and KiDS datasets. However, instead of relying on direct model outputs, CLAP leverages known redshifts to construct and calibrate probability density estimates, thus it has a better ability in reducing the mean redshift biases. While \citet{Lin2022} also tackled the mean redshift biases and potentially the impact of miscalibration, the accuracy might be compromised for bias correction. This is illustrated by the conspicuous boost in $\sigma_{\mathrm{MAD}}$ and the strong tilde in $\Delta F_{h-h}$ for the CFHTLS dataset. In contrast, this trade-off is no longer a problem for CLAP, thus CLAP achieves high accuracy comparable to other image-based methods.

Furthermore, we point out that the results produced by the image-based methods but the ones from \citet{Lin2022} are all shown in a simple scenario in which the distribution for the target sample matches that for the training sample, giving an impression that CLAP does not noticeably outperform the other conventional methods especially for the SDSS dataset. In general scenarios, however, the mean redshift residuals obtained by the other methods would deviate from zero if shown for a target sample with a different distribution, as the distribution learned from the training sample is fixed in the trained model. On the contrary, CLAP is flexible to adjust the learned distribution, outperforming the other methods in resolving the mean redshift biases induced by mismatches. The next paper on CLAP will discuss how to estimate an unknown target distribution and resolve mismatches based on the CLAP implementation.

Despite the merits, CLAP inevitably suffers from non-negligible errors at high redshift owing to low S/N and insufficient spectroscopic data, the same as other deep learning methods. The errors at high redshift might be the cause for uncontrolled mean redshift biases such as the drop of $\delta_{<z>}$ at $r > 22$ for the CFHTLS dataset and the mild decreasing trend as a function of $r$-band magnitude for the KiDS dataset. We suggest that a good control of errors is crucial for ensuring the robustness of CLAP. This would require a larger collection of good-quality spectroscopic data at the high-redshift regime, which are expected to be available from future spectroscopic surveys. Furthermore, in order to optimise the exploitation of available data, it is promising to extend CLAP by exploring the hybridisation of complementary methods \citep[e.g.][]{Sanchez2018, Alarcon2020, Rau2022} as well as domain knowledge-motivated algorithms, which would bring in improvements in high redshift estimation.

\begin{figure*}
\centering
\includegraphics[width=0.95\linewidth]{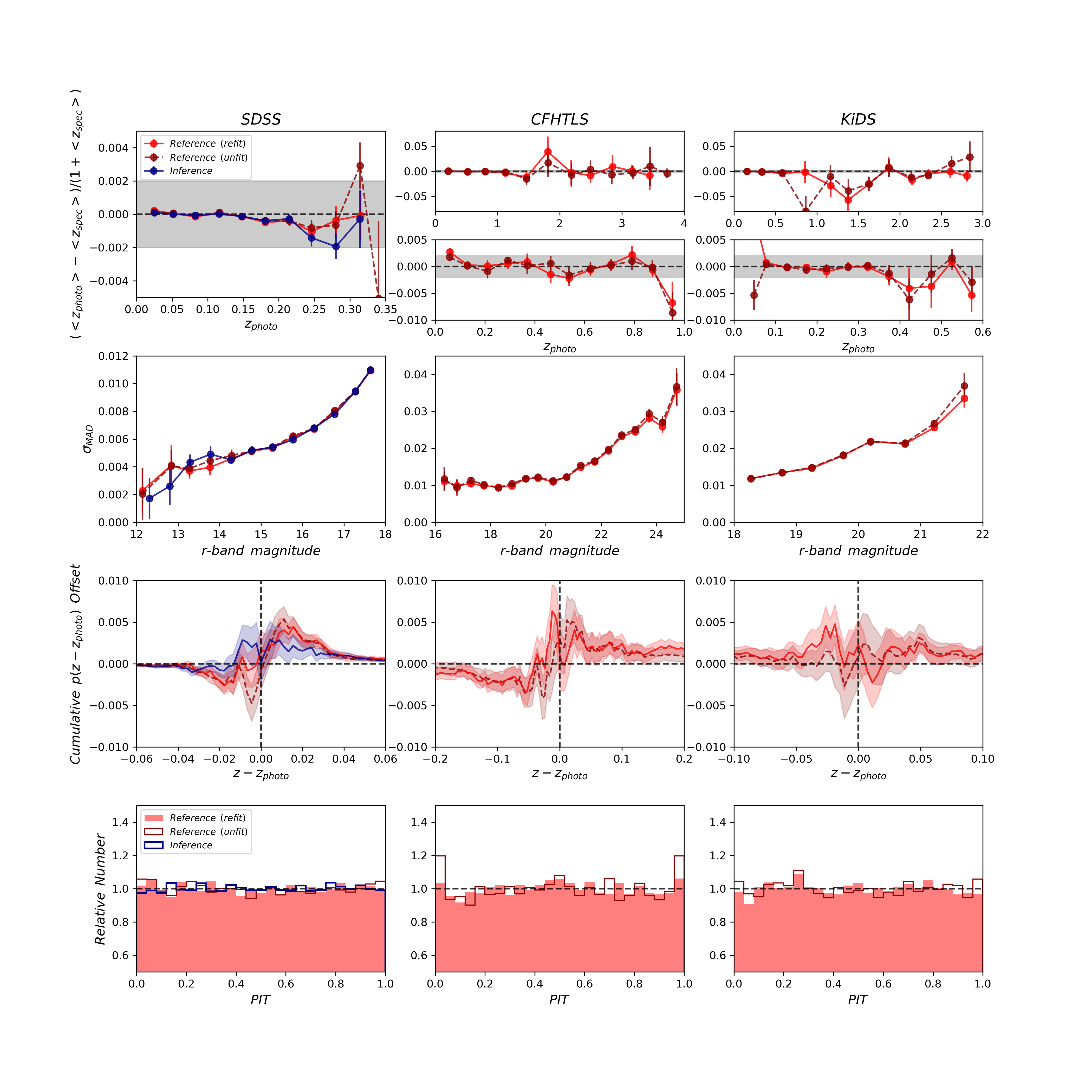}
\caption{Outcomes of resuming end-to-end discriminative models by refitting in CLAP illustrated for the SDSS, CFHTLS, and KiDS target samples. The SDSS target sample is split into a reference subsample (used to refit the model) and an inference subsample (held out for inference), while the CFHTLS and KiDS target samples are only used as reference samples. The results obtained by No-NIR CLAP for the KiDS dataset are not shown in order to avoid redundancy. For each dataset, only one of the ten CLAP models in the ensemble is selected for demonstration. \textit{First row:} Mean redshift residuals as a function of photometric redshift $z_{photo}$ computed using the reference samples with and without refitting. The results for the SDSS inference subsample are also shown. As in Fig.~\ref{fig:main_point_est}, the results for the CFHTLS and KiDS datasets are shown in the low redshift ranges in addition to the full ranges. \textit{Second row:} $\sigma_{\mathrm{MAD}}$ as a function of $r$-band magnitude. \textit{Third row:} Cumulative offsets $\Delta F_{p_{\sum}-h}$ between each of the stacked recentred probability densities $p_{\sum}(z - z_{photo})$ and the corresponding $z_{spec} - z_{photo}$ histogram. The vertical dashed line in each panel indicates the zero point where $z$ and $z_{photo}$ coincide. \textit{Fourth row:} PIT distributions. The $y$-axis in each panel is truncated at 0.5 for clearer illustration.}
\label{fig:refit_pointest_pit}
\end{figure*}

\subsection{Resumption of end-to-end models by refitting} \label{sec:assess_refitting}

Refitting has been implemented in CLAP to resume end-to-end discriminative models in order to avoid running computationally expensive KNN for large-scale imaging data. In Fig.~\ref{fig:refit_pointest_pit}, we illustrate the outcomes of refitting for the SDSS, CFHTLS, and KiDS target samples that are used as reference samples, and also the results for the SDSS inference subsample. The results obtained by No-NIR CLAP for the KiDS dataset are similar to those obtained by default CLAP and are thus not shown in order to avoid redundancy. With no loss of generality, for each dataset we only selected one of the ten CLAP models in the ensemble for demonstration. It can be seen that the mean redshift residuals $\delta_{<z>}$, $\sigma_{\mathrm{MAD}}$, the cumulative offsets $\Delta F_{p_{\sum}-h}$, and the PIT distributions for each reference sample with and without refitting and for the SDSS inference subsample are all in good consistency. In particular, the deviations in the cumulative offsets $\Delta F_{p_{\sum}-h}$ are considered insignificant since they are no more than the upper levels set for each dataset in Fig.~\ref{fig:main_pdfsc} (i.e. 0.005, 0.01, and 0.01, respectively). The PIT distributions remain nearly flat regardless of refitting or inference. 

We thus confirm that no apparent biases were introduced in our refitting procedure or the inference process. The resumption of end-to-end models by refitting holds promises as it not only preserves the advantages of conventional deep learning methods (i.e. accuracy and efficiency), but also retains all the gains from KNN. In addition, the smoothing applied on the probability density estimates effectively remove the unrealistic spikes due to insufficient data coverage (see Appendix~\ref{sec:examples}).

\section{Investigation on miscalibration} \label{sec:miscalibration}

\begin{figure*}
\centering
\includegraphics[width=1.0\linewidth]{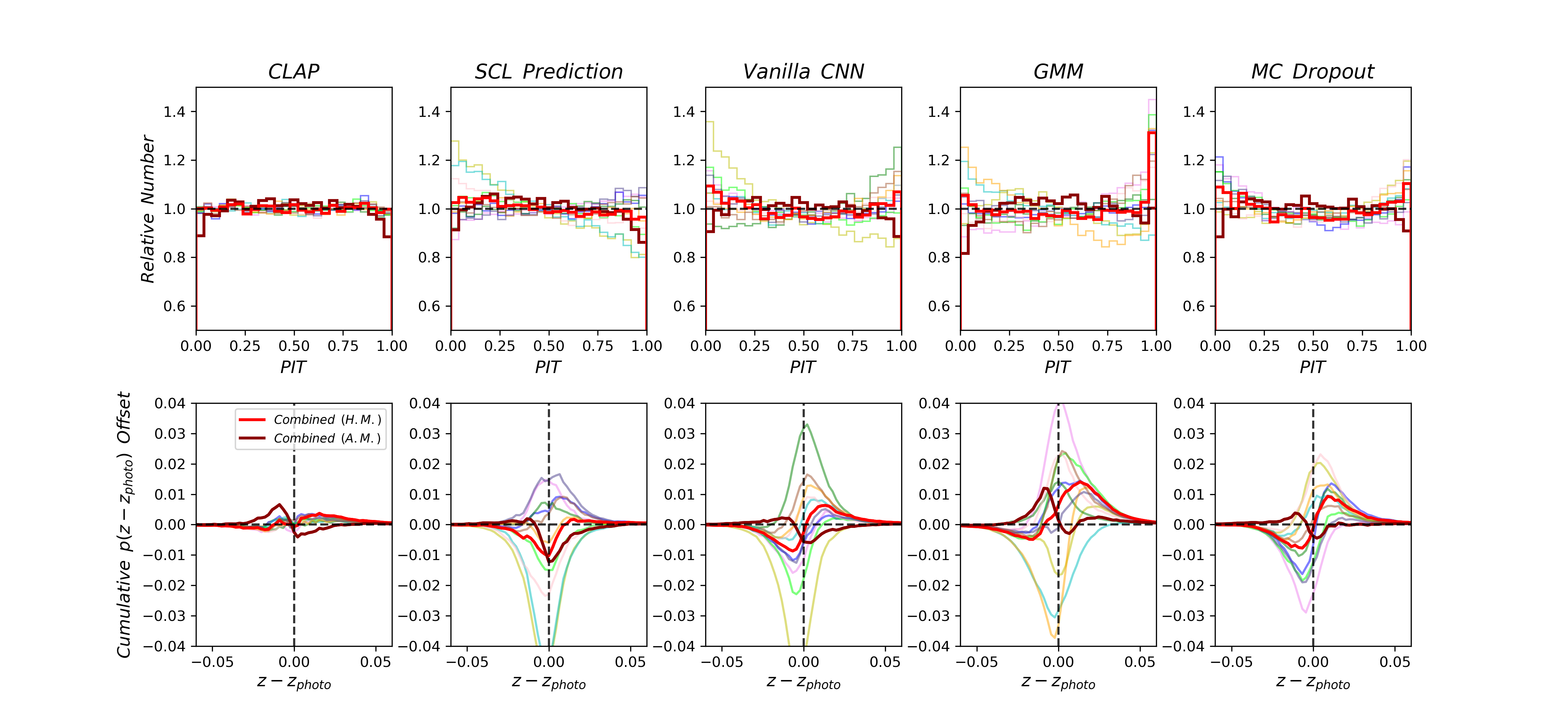}
\caption{PIT distributions and cumulative offsets $\Delta F_{p_{\sum}-h}$ for CLAP, the SCL prediction, the vanilla CNN, the GMM, and the MC dropout (detailed in the text). The results for the ten uncombined probability density estimates in the ensemble for each method are shown in different colours. The results obtained by combining the ten estimates using the harmonic mean (H.M.) are shown in red, while those obtained using the arithmetic mean (A.M.) are shown in dark red. For the PIT, the $y$-axis in each panel is truncated at 0.5 for clearer illustration. For the cumulative offsets, the vertical dashed line in each panel indicates the zero point where $z$ and $z_{photo}$ coincide.}
\label{fig:example_miscalibration}
\end{figure*}

\begin{figure}
\centering
\includegraphics[width=1.0\linewidth]{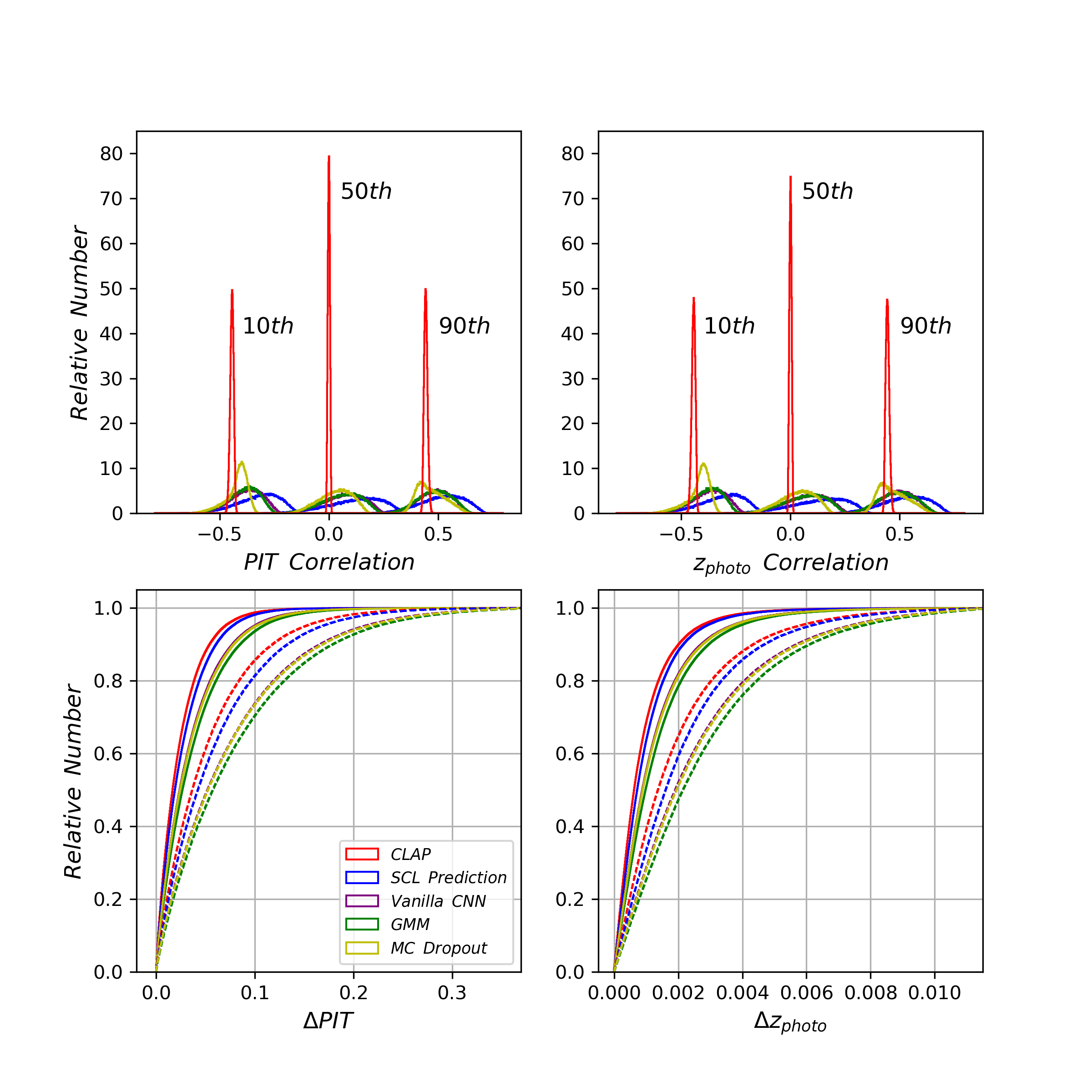}
\caption{Correlation and uncertainty analysis for CLAP, the SCL prediction, the vanilla CNN, the GMM, and the MC dropout (detailed in the text). \textit{Upper panels:} 10th, 50th, and 90th percentiles drawn from the distributions of PIT and $z_{photo}$ correlations between each data instance (viewed as a ten-dimensional vector) and all the other data instances from the sample. \textit{Lower panels:} Distributions of PIT and $z_{photo}$ uncertainties that are not accounted for by uncombined probability density estimates (dashed curves) or by a combination of five estimates (solid curves).}
\label{fig:corr_delta_miscalibration}
\end{figure}

Miscalibration is a problem that jeopardises reliable probability density estimation. We attempted to illustrate and analyse miscalibration in the context of deep learning-based photometric redshift estimation with CLAP as a reference. Without losing generality, this analysis was conducted using the SDSS target sample alone. The reference and the inference subsamples are not separated.

\subsection{Illustration of miscalibration} \label{sec:mis_illustration}

We illustrated miscalibration via a comparison among CLAP, the SCL prediction, and a few representative conventional deep learning methods, including:
\begin{itemize}
    \item a vanilla CNN in which we kept the encoder and the estimator from CLAP but without implementing the SCL scheme, trained using the one-hot spectroscopic redshift labels and the softmax cross-entropy loss function only, similar to \citet{Pasquet2019};
    \item a network same as the vanilla CNN, but the output layer was replaced by a mixture of five Gaussian components (i.e. GMM), and trained using the one-hot redshift labels and the CRPS loss function, following the implementation by \citet{DP2018};
    \item a network same as the vanilla CNN, but Monte Carlo (MC) dropout \citep{GG2016} was applied on two fully connected layers with a dropout rate of 0.5, and the harmonic means of MC estimates (i.e. softmax outputs) were taken as the final probability density estimates, which is an implementation of BNN.
\end{itemize}
For each of these methods, we trained ten models using the same number of training iterations, mini-batch size, and learning rate as used in SCL for implementing CLAP. 

In Fig.~\ref{fig:example_miscalibration}, we show the PIT distribution and the cumulative offset $\Delta F_{p_{\sum}-h}$ for each of the ten (uncombined) models obtained by each method. Same as those shown in Fig.~\ref{fig:main_pdfsc}, the cumulative offset $\Delta F_{p_{\sum}-h}$ characterises the normalised deviation of the stacked recentred probability density $p_{\sum}(z - z_{photo})$ relative to the corresponding $z_{spec} - z_{photo}$ histogram. It can be clearly seen that CLAP provides robust results for all of the ten models in the ensemble, whereas the other methods all produce non-negligible deviations from uniform PIT distributions and zero offsets. These deviations typically have large variations among the ten realisations for each of the methods, exhibiting different degrees of miscalibration, despite that the models for each method were developed in the same manner. The variable deviations may originate from different sources of stochasticity introduced in the training process such as random initialisation and random selections of mini-batches, on top of the impacts of overfitting and data imbalance. They offer an evident indication that the probability density estimates produced by the conventional methods are prone to miscalibration \citep[see also][]{Guo2017, Thulasidasan2019, Minderer2021, Wen2021}. Furthermore, Fig.~\ref{fig:example_miscalibration} shows the outcomes of combining the ensemble of ten estimates using the harmonic mean (H.M.) or the arithmetic mean (A.M.). With this, we demonstrate that properly combining probability density estimates (i.e. using the harmonic mean) would lessen the degree of miscalibration, yet it cannot guarantee a complete removal. In contrast, the estimates from CLAP do not suffer from apparent miscalibration either before or after combination using the harmonic mean.

A deeper understanding of these findings is discussed in the next subsection. Additionally, remarks on combining probability density estimates using the harmonic mean or the arithmetic mean are given in Sect.~\ref{sec:HMvsAM}.

\subsection{Association with uncertainties and correlations} \label{sec:mis_association}

Other than the dispersions intrinsic to the mapping between photometric data and redshift, a reliable probability density should typically encompass two types of uncertainties: `aleatoric' uncertainties due to intrinsic stochastic noise in data and `epistemic' uncertainties due to incomplete knowledge or limitations of the method implementation. There are several components in epistemic uncertainties. One component reflects the incapability or limit of an estimation method that has a certain configuration (e.g. the network architecture, the learning rate, the number of training iterations). For example, a shallow network would not be as capable as a deep network in extracting redshift information from galaxy images, thus the shallow network would produce larger uncertainties. We anticipate that both this component of epistemic uncertainties and aleatoric uncertainties are encapsulated in uncombined probability density estimates from CLAP given sufficient statistics, as both the errors due to model incapability and the errors in data are propagated to the latent space in the training process and taken into account by KNN. There are other components of epistemic uncertainties that stem from stochasticity in the method implementation, such as a particular random initialisation, variability in the iterative training procedure, and particular overfitted features. We argue that these components are often not accounted for by uncombined probability density estimates from a single measurement. This leads to the consequence that the probability densities produced by two models with the same network architecture but different parametrisations may not be identical. Nonetheless, we expect these unaccounted-for uncertainties to be reduced by properly combining different measurements.

On the other side, the self-adaptive training procedure of a deep learning model would correlate its outputs for different data instances, as the outputs are conditioned on the same network with a certain parametrisation. This would induce an excess of correlated errors in bulk that cannot be cancelled out and may be problematic for downstream astrophysical and cosmological analysis.

In this subsection, we attempted to investigate how miscalibration is associated with both unaccounted-for uncertainties and correlations between data instances induced by the estimation method. We first analysed the correlations of PIT and $z_{photo}$ estimated using the ensemble of probability density estimates. With each quantity, every data instance can be regarded as a ten-dimensional vector. We computed the correlation between every two data instances over the whole sample. Then for each data instance, we drew three percentiles (i.e. 10th, 50th, 90th) from the distribution of correlations with all the other instances.

In the upper panels of Fig.~\ref{fig:corr_delta_miscalibration}, we show the distributions of the three correlation percentiles for PIT and $z_{photo}$, comparing CLAP and the other methods listed in Sect.~\ref{sec:mis_illustration}. A sharp contrast is noteworthy: the percentile distributions for CLAP are all narrow spikes, with the 50th percentiles located precisely at zero, whereas the distributions for the other methods are much more dispersed and favour positive correlations (though the use of MC dropout may reduce the dispersions to some extent). This contrast marks an excess of method-induced correlations for the conventional methods. The existence of such excessive correlations implies that there would be bulk shifts in response to perturbing the measurement for any particular data instance that has non-zero net correlations with the whole sample, resulting in correlated miscalibration behaviours among data instances and aggravating the variations of the non-uniform PIT distributions and the non-zero cumulative offsets $\Delta F_{p_{\sum}-h}$ as manifested in Fig.~\ref{fig:example_miscalibration}. In other words, miscalibration is sensitive to the excessive correlations. They may be produced because the one-hot labels used for training these conventional methods are devoid of power in constraining the actual shapes of the probability densities, leaving the freedom for generating correlated errors in bulk. On the contrary, CLAP adopts KNN-calibrated labels for refitting that have much better constraining power and strongly suppress correlations induced by the self-adaptive training procedure. The resulting correlation distributions are thus symmetric and steady for the whole sample, and no significant non-zero net correlations are present.

Next, we analysed the uncertainties of PIT and $z_{photo}$. For each data instance, we randomly selected two measurements out of the ensemble and computed their absolute difference. The distribution of the differences over the sample can be viewed as a manifestation of the typical epistemic uncertainties not accounted for by uncombined probability density estimates. Randomly selecting estimates from the ensemble can average the contributions from correlated errors between data instances. Similarly, for each data instance, we computed the absolute difference between two measurements, one obtained by combining five randomly selected probability density estimates using the harmonic mean while the other by combining the remaining five estimates, which would manifest the unaccounted-for epistemic uncertainties for a combination of five estimates.

The lower panels of Fig.~\ref{fig:corr_delta_miscalibration} present the cumulative distributions for the uncombined and combined cases. We confirmed that the unaccounted-for uncertainties can indeed be reduced by combining the ensemble of estimates, which in turn lowers the degree of miscalibration for the other methods. For CLAP, the unaccounted-for uncertainties approximately shrink by two times when five estimates are combined. We found that the unaccounted-for uncertainties mildly evolve with $r$-band magnitude, as there tend to be large estimation errors for high-magnitude galaxies. Nonetheless, by combining ten estimates as in our main experiments, we anticipate that the unaccounted-for uncertainties can be further reduced and reach $\Delta z_{photo} \sim 0.001$ for individual data instances at the 90th percentile for the SDSS sample.

Furthermore, there are larger uncertainties for the other methods compared to CLAP, roughly 10\%, 40\%, 50\%, and 40\% larger for the SCL prediction, the vanilla CNN, the GMM, and the MC dropout, respectively. These ratios remain nearly unchanged when the overall uncertainties are reduced. This indicates that the excessive uncertainties may be contributed from the excessive method-induced correlations. These correlations remain approximately constant such that the excessive uncertainties scale with the overall uncertainties. As a result, the apparent miscalibration may be alleviated but it is hard to be eliminated for the other methods provided that a reduction of uncertainties does not lead to a reduction of correlations. For the SCL prediction in particular, while a 10\% excess of uncertainties seems insignificant, the presence of excessive correlations still results in obvious miscalibration behaviours that cannot be removed completely by combining the ensemble of estimates, as already shown in Fig.~\ref{fig:example_miscalibration}.

To summarise, both unaccounted-for uncertainties and excessive correlations induced by the method would exacerbate miscalibration. We speculate that the excessive correlations may stem from the use of one-hot labels that lack constraining power. They would stay unchanged when the uncertainties are reduced, and cause correlated miscalibration behaviours and contribute to the overall unaccounted-for uncertainties. Owing to their existence, reducing the uncertainties may not guarantee to remove miscalibration. Unlike the conventional methods, CLAP does not suffer from the issue of excessive correlations or apparent miscalibration.

Hence, for developing deep learning methods to obtain well-calibrated probability density estimates suitable for astrophysical and cosmological applications, we suggest that one has to pay attention to the interplay between miscalibration, unaccounted-for uncertainties, and method-induced correlations, with a particular focus on the negative impact of correlated errors. In principle, miscalibration may be reduced by, for example, carefully fine-tuning a discriminative model with an optimised training procedure and checking the outputs using a validation sample, yet we note that this may be difficult to effectively circumvent the issue of excessive correlations. In this regard, we suggest that it would be favourable to replace one-hot labels with more informative labels that have better access to the actual shapes of probability densities for training deep learning models. Such informative labels may be constructed using KNN-based methods such as CLAP, or extended with complementary approaches that are independent of the exploitation of photometric data, such as spatial cross-correlation for individual galaxies \citep[e.g.][]{Rahman2015, Morrison2017, Sanchez2018, Scottez2018}. Furthermore, using informative labels bypasses the need for a particular fine-tuning procedure.

\subsection{Harmonic mean versus arithmetic mean for combining probability density estimates} \label{sec:HMvsAM}

Finally in this subsection, we give some remarks on combining an ensemble of probability density estimates using the harmonic mean or the arithmetic mean, as it also concerns the miscalibration issue. Conceptually speaking, there are specific errors introduced by each model $\Phi$ from the ensemble, which are propagated to the learned joint distribution $p(z, \Phi(d))$ of redshifts and the projections of data. If combining probability density estimates using the arithmetic mean as suggested by \citet{Pasquet2019}, $\Phi(d)$ is treated as a fixed quantity; the specific errors from all the models to be combined would not be removed but projected and aggregated along the $z$ dimension. Therefore, using the arithmetic mean would artificially increase epistemic uncertainties, broaden the probability density estimates, and introduce miscalibration, though the unaccounted-for uncertainties can be reduced. This has been illustrated in Fig.~\ref{fig:example_miscalibration}. However, we suggest that using the harmonic mean would essentially project the specific errors on the $\Phi(d)$ dimension in each $z$ bin. Then the errors in any $z$ bins can be reduced by averaging the probabilities $p(\Phi(d))$ under the first-order approximation, as proved in Sect.~\ref{sec:combine}. It turns out that no miscalibration appears as a result of using the harmonic mean, thus it deserves to be a proper way to combine probability density estimates. As combining non-identical estimates is a non-trivial task, while other approaches are possible to use \citep[e.g.][]{Dahlen2013}, care must be taken to ensure statistical soundness and avoid violating the frequentist definition.

\section{Conclusion} \label{sec:conclusion}

We developed a novel method called the \textbf{C}ontrastive \textbf{L}earning and \textbf{A}daptive KNN for \textbf{P}hotometric Redshift (CLAP), which empowers state-of-the-art image-based deep learning methods with the advantages of $k$-nearest neighbours (KNN) for photometric redshift probability density estimation. With CLAP, we have resolved the miscalibration encountered by conventional deep learning, the inconsistency between accuracy and confidence that leads to a disagreement between the estimated and the empirical probability densities. This is crucial for obtaining probability density estimates that are required to be well-calibrated and suitable for downstream astrophysical and cosmological inference.

The main idea of CLAP is to transform a conventional discriminative model into a supervised contrastive learning (SCL) framework that establishes a latent space encoding redshift information, and then KNN can be applied to substitute the conventional uncalibrated network output (e.g. softmax) for probability density estimation. The deep learning-based contrastive learning is combined with supervised learning using spectroscopic redshifts as labels, so that meaningful redshift information can be extracted from images and propagate to the latent space, enabling CLAP to achieve much higher accuracy than by photometry-only KNN approaches. The adaptive KNN algorithm and the KNN-enabled recalibration procedure rely on diagnostics based on the local distributions of the probability integral transform (PIT) values, which naturally ensures that each probability density estimate is locally calibrated.

We applied CLAP to the SDSS, CFHTLS, and KiDS datasets that have different redshift coverages (i.e. $z_{spec}<0.4$, $z_{spec}<4.0$, and $z_{spec}<3.0$, respectively). To evaluate the obtained probability density estimates according to the frequentist definition, the probability-weighted mean redshifts were taken as point estimates. Compared to benchmark methods, we found that CLAP is indeed capable of giving smaller distribution offsets (i.e. between the stacked recentred probability density $p_{\sum}(z - z_{photo})$ and the $z_{spec} - z_{photo}$ histogram), producing better PIT distributions, and having a better control of mean redshift biases. Our experiments also verified that CLAP does not suffer from the bias-variance trade-off faced by \citet{Lin2022}. It achieves high accuracy comparable to that by conventional image-based deep learning methods.

Furthermore, we demonstrated that the resumption of end-to-end discriminative models via refitting retains the benefits from KNN, but bypasses its expensive computation for processing large-scale data, regaining the computational efficiency of deep learning. Therefore, we confirmed that CLAP can leverage the advantages of both deep learning and KNN at the same time. For practical purposes, we suggest that such resumed models can be used to make inference for a large target sample, while the computationally intensive KNN may only be needed to obtain labels for a small representative subset used for refitting.

With reference to CLAP, we investigated the miscalibration issue for conventional deep learning discriminative models. We experimented on a few representative conventional methods, and found that they all exhibit various degrees of miscalibration with large variations among different realisations. For these methods, we note that miscalibration is particularly sensitive to an excess of method-induced correlations among data instances other than unaccounted-for epistemic uncertainties, which may be due to the use of one-hot labels that do not have access to the actual shapes of probability densities. On the contrary, CLAP is not affected by excessive correlations or apparent miscalibration because the probability density estimates are directly calibrated using the nearest neighbours whose spectroscopic redshifts are known.

We also confirmed that combining the probability density estimates produced by an ensemble of realisations can indeed reduce unaccounted-for epistemic uncertainties and improve accuracy. However, a proper way of combining probability density estimates should be applied in order to avoid artificially introducing epistemic uncertainties and miscalibration. Contrary to the seemingly intuitive arithmetic mean, we proved that the harmonic mean should be considered as a better choice.

In summary, by efficaciously resolving miscalibration in the context of photometric redshift estimation, our method CLAP has made a promising step towards developing robust and reliable deep learning methods for astrophysical problems. In the next paper in our series, we will leverage CLAP to tackle the mismatch issue, another common but challenging shortcoming of conventional deep learning, and more broadly, data driven methods. In general, we anticipate that hybrid methods may be developed and integrated with domain knowledge to provide complementary information for photometric redshift estimation and extend our current implementation of CLAP. This is essential for alleviating the difficulty in controlling errors and biases at high redshift due to low S/N and insufficient spectroscopic data. We also expect CLAP to be used in applications that involve probability density estimation not limited to photometric redshift. The extension of CLAP is left for future endeavours.

\section*{Data and Code Availability} \label{sec:data_code_avail}

The photometric redshift catalogues produced by default CLAP are available at \url{https://zenodo.org/records/13954481}.
The code used in this work is available at \url{https://github.com/QiufanLin/CLAP-I}.

\begin{acknowledgements}

This work is supported by `The Major Key Project of PCL' and makes use of the Cloud Brain System at PCL for computing resources.

H.X.R. acknowledges financial support from the National Natural Science Foundation of China (62201306).

Y.S.T. acknowledges financial support from the Australian Research Council through DECRA Fellowship DE220101520.

The authors thank St\'{e}phane Arnouts and Marie Treyer for access to the CFHTLS imaging data and catalogues.

This work makes use of Sloan Digital Sky Survey (SDSS) data. Funding for SDSS-III has been provided by the Alfred P. Sloan Foundation, the Participating Institutions, the National Science Foundation, and the U.S. Department of Energy Office of Science. The SDSS-III web site is http://www.sdss3.org/. SDSS-III is managed by the Astrophysical Research Consortium for the Participating Institutions of the SDSS-III Collaboration including the University of Arizona, the Brazilian Participation Group, Brookhaven National Laboratory, Carnegie Mellon University, University of Florida, the French Participation Group, the German Participation Group, Harvard University, the Instituto de Astrofisica de Canarias, the Michigan State/Notre Dame/JINA Participation Group, Johns Hopkins University, Lawrence Berkeley National Laboratory, Max Planck Institute for Astrophysics, Max Planck Institute for Extraterrestrial Physics, New Mexico State University, New York University, Ohio State University, Pennsylvania State University, University of Portsmouth, Princeton University, the Spanish Participation Group, University of Tokyo, University of Utah, Vanderbilt University, University of Virginia, University of Washington, and Yale University.

This work is based on observations obtained with MegaPrime/MegaCam, a joint project of CFHT and CEA/DAPNIA, at the Canada-France-Hawaii Telescope (CFHT) which is operated by the National Research Council (NRC) of Canada, the Institut National des Sciences de l'Univers of the Centre National de la Recherche Scientifique (CNRS) of France, and the University of Hawaii. This work is based in part on data products produced at Terapix and the Canadian Astronomy Data Centre as part of the Canada-France-Hawaii Telescope Legacy Survey, a collaborative project of NRC and CNRS.

This work is based on observations made with ESO Telescopes at the La Silla Paranal Observatory under programme IDs 177.A-3016, 177.A-3017, 177.A-3018 and 179.A-2004, and on data products produced by the KiDS consortium. The KiDS production team acknowledges support from: Deutsche Forschungsgemeinschaft, ERC, NOVA and NWO-M grants; Target; the University of Padova, and the University Federico II (Naples).

\end{acknowledgements}

\bibliographystyle{aa}
\bibliography{pz_pdf}

\begin{appendix}

\section{Network architectures} \label{sec:network}

\begin{figure*}
\centering
\centerline{\includegraphics[width=1.1\linewidth]{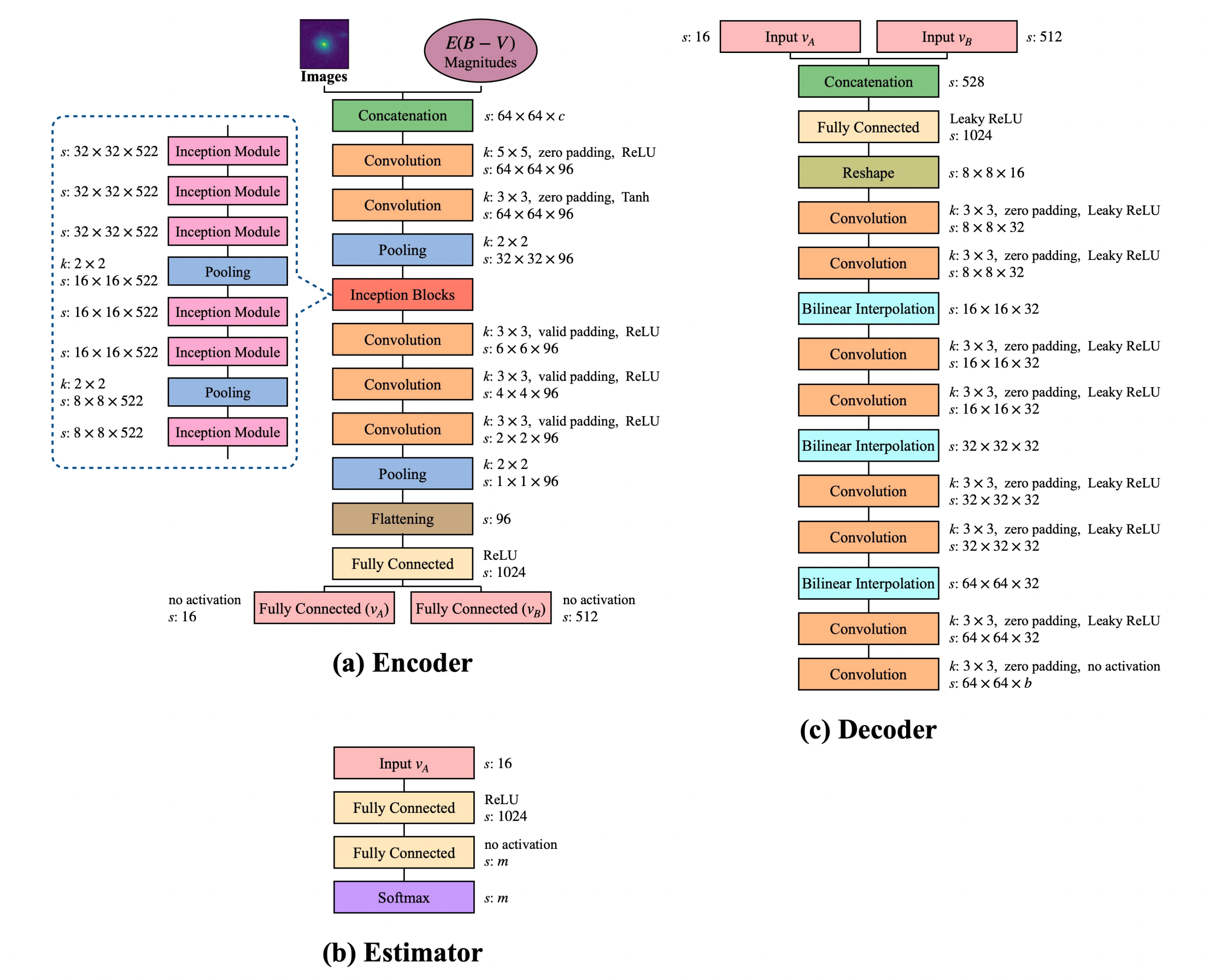}}
\caption{Network architectures of the encoder, the estimator, and the decoder used in our work. The kernel size, the padding type, the activation function, and the output shape for each layer are specified when necessary. $k$ and $s$ refer to the kernel size and the output shape, respectively. In the encoder, $c$ stands for the total number of channels after concatenating multi-band images and additional input ($c=6$ for the SDSS dataset and the CFHTLS dataset; $c=10$ for the KiDS dataset). The encoder outputs the latent vector $v_A$ and the vector $v_B$ with parallel fully connected layers. In the decoder, $b$ stands for the number of optical bands for images ($b=5$ for the SDSS dataset and the CFHTLS dataset; $b=4$ for the KiDS dataset). The Leaky ReLU activation has a leaky ratio of 0.2. In the estimator, $m$ stands for the number of redshift bins in the output (quoted in Table~\ref{tab:coverage}).}
\label{fig:architecture}
\end{figure*}

We adopted a modified version of the inception network developed by \citet{Treyer2024} for the encoder network and the estimator network in our SCL framework. The detailed description of the inception network is presented therein. We designed a decoder that leverages bilinear interpolation for upsampling images. The architectures of the encoder, the estimator, and the decoder are presented in Fig.~\ref{fig:architecture}.

\section{Examples of probability density estimates} \label{sec:examples}

\begin{figure*}
\centering
\includegraphics[width=1.0\linewidth]{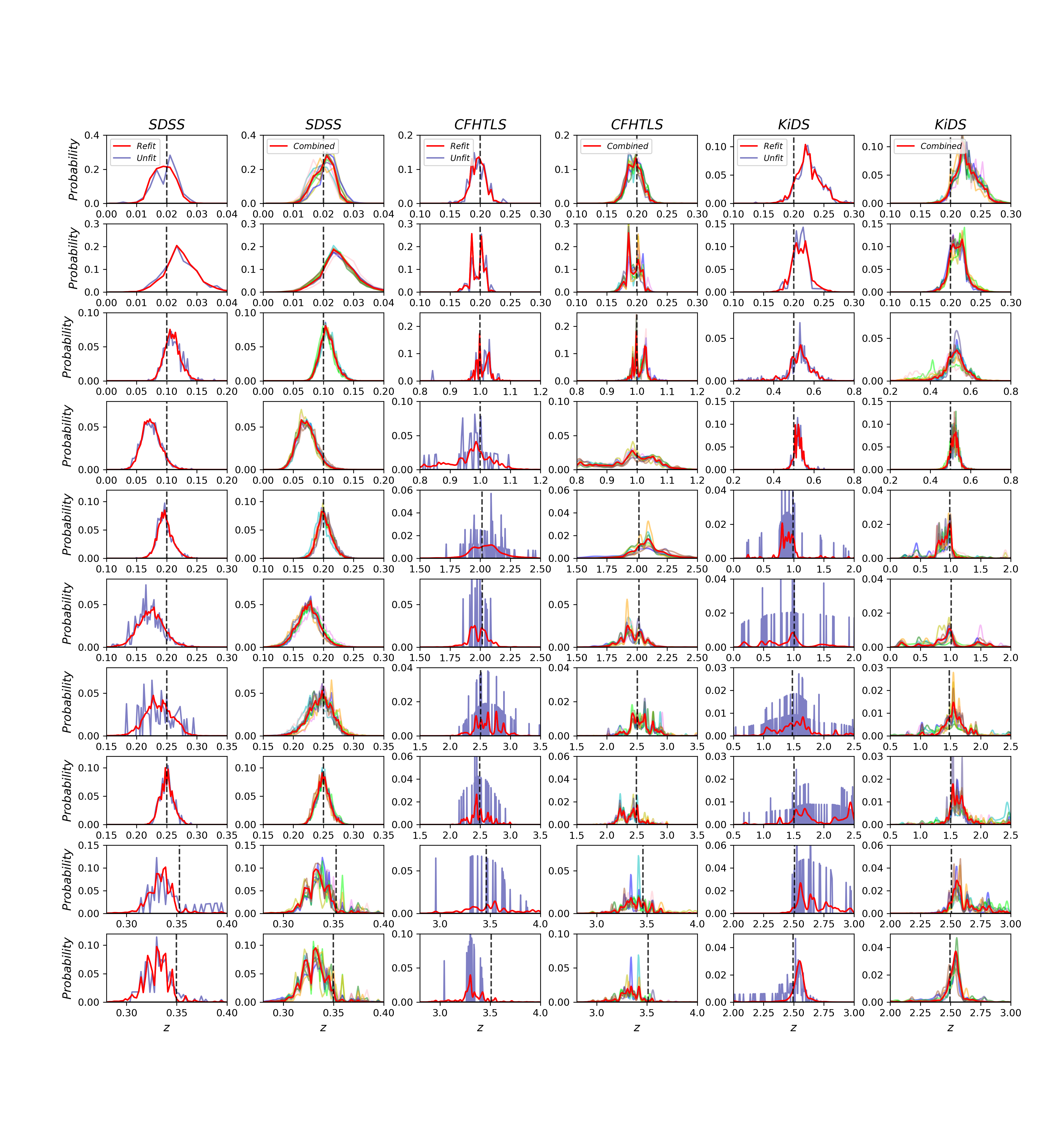}
\caption{Examples of probability density estimates obtained by CLAP from the SDSS, CFHTLS, and KiDS datasets. \textit{First column:} Examples of uncombined probability density estimates for galaxies at different redshifts from the SDSS dataset, randomly selected from one of the ten CLAP models in the ensemble. Comparison is made for the same galaxies with and without refitting. In each panel, the vertical dashed line indicates the spectroscopic redshift. \textit{Second column:} Probability density estimates after refitting for the same galaxies in the first row. The ten uncombined estimates are shown in different colours, while the ones obtained by combining the ten estimates using the harmonic mean are shown in red. \textit{Third and fourth columns:} Same as the first two rows, but from the CFHTLS dataset. \textit{Fifth and sixth columns:} Same as the first two rows, but from the KiDS dataset.}
\label{fig:example_pdfs}
\end{figure*}

We present the probability density estimates obtained by CLAP for a few exemplar galaxies at different redshifts in Fig.~\ref{fig:example_pdfs}. We illustrate the outcomes with and without applying the refitting procedure, and before and after combining the ensemble using the harmonic mean. As shown, refitting preserves the global shape of each probability density estimate. There are many spikes in the raw estimates when the data are insufficient to populate the redshift bins, especially for the CFHTLS and KiDS datasets at high redshift. These unrealistic features can be effectively removed by the smoothing applied in refitting. There are non-negligible variations among the uncombined estimates due to unaccounted-for epistemic uncertainties. The combination using the harmonic mean takes into account the variations but does not aggregate the uncertainties. As demonstrated in the main text, both refitting and combining estimates using the harmonic mean do not introduce apparent biases or miscalibration.

\section{Assessment on recalibration} \label{sec:assess_recalibration}

\begin{figure*}
\centering
\includegraphics[width=1.0\linewidth]{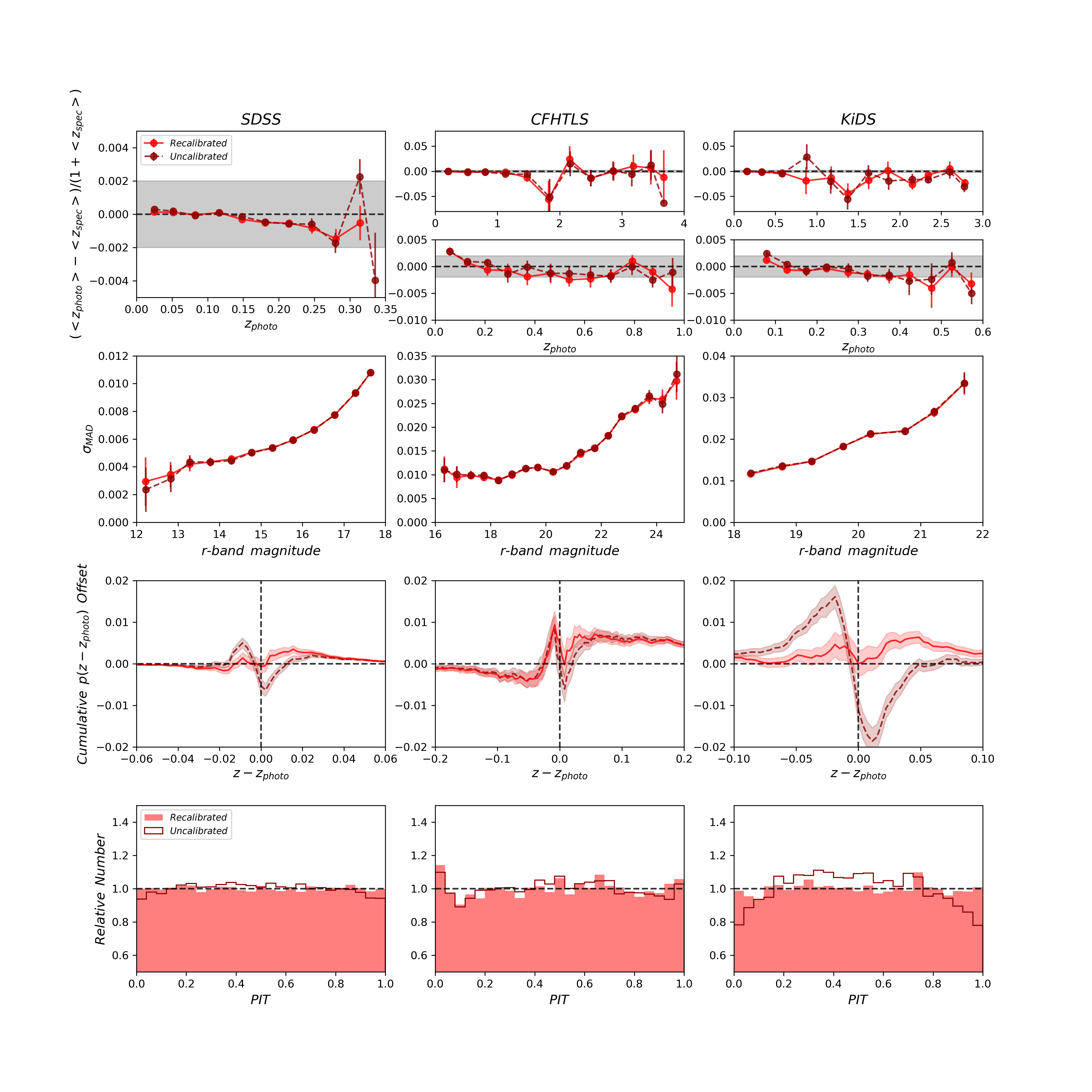}
\caption{Same as Fig.~\ref{fig:refit_pointest_pit}, but showing the outcomes with and without recalibration implemented in CLAP. For each dataset, the ensemble of ten probability density estimates are combined using the harmonic mean.}
\label{fig:recal_pointest_pdfsc_pit}
\end{figure*}

The recalibration procedure is a crucial part in CLAP as it ensures good calibration for each data instance. We demonstrate its importance in Fig.~\ref{fig:recal_pointest_pdfsc_pit}, where we compare the results with and without recalibration for the SDSS, CFHTLS, and KiDS datasets. These results are obtained by combining the ensemble of ten probability density estimates using the harmonic mean. As shown, although there seems to be no conspicuous difference with and without recalibration in terms of the point estimates $z_{photo}$, the recalibration procedure indeed improves the calibration of the obtained probability density estimates. Particularly for the KiDS dataset, the probability density estimates without recalibration would appear to be overdispersed, as indicated by the non-negligible cumulative offsets $\Delta F_{p_{\sum}-h}$ and the concave PIT distribution, while the estimates become much better calibrated after recalibration. Furthermore, we note that for over 90\% of the probability density estimates, the required correction by recalibration is mild, equivalent to an amount of no more than a redshift bin $\Delta z$. This implies that the adaptive KNN algorithm may occasionally produce a biased $k$ value that does not define a neighbourhood properly, yet the local recalibration is a remedy for this bias.

\section{Statistics for point estimates} \label{sec:statistics_point}

We present the global mean residual $\left< \delta z \right> = \left< (z_{photo} - z_{spec})/(1 + z_{spec}) \right>$, $\sigma_{\mathrm{MAD}}$, and the outlier fraction $\eta_{>\alpha}$ with $|\delta z|>\alpha$ for the photometric redshift point estimates obtained by all the methods presented in Fig.~\ref{fig:main_point_est}. The results for the SDSS, CFHTLS, and KiDS datasets are quoted in Tables~\ref{tab:stats_SDSS}, \ref{tab:stats_CFHTLS}, and \ref{tab:stats_KiDS}, respectively. $\alpha=0.05$ is adopted for the SDSS dataset, while $\alpha=0.15$ is for the CFHTLS dataset and the KiDS dataset.

\begin{table}\footnotesize
\centering
\begin{tabular}{l | c c c}
\hline
   &  $<\delta z>$  &  $\sigma_{\mathrm{MAD}}$  &  $\eta_{>0.05}$  \\
\hline
CLAP  &  0.00005  &  0.00853  &  $0.20\%$  \\
\hline
SCL Prediction  &  0.00024  &  0.00850  &  $0.21\%$  \\
\hline
Inception Net + Bias Corr. \citep{Lin2022}  &  0.00002  &  0.00919  &  $0.41\%$  \\
\hline
Inception Net \citep{Pasquet2019}  &  0.00009  &  0.00915  &  $0.31\%$  \\
\hline
Capsule Net \citep{Dey2022}  &  0.00007  &  0.00898  &  $0.19\%$  \\
\hline
KNN + Regression \citep{Beck2016}  &  0.00064  &  0.01342  &  $1.36\%$  \\
\hline
\end{tabular}
\caption{Statistics for the point estimates obtained by the methods presented in Sect.~\ref{sec:results} for the SDSS dataset.} \label{tab:stats_SDSS}
\end{table}

\begin{table}\footnotesize
\centering
\begin{tabular}{l | c c c}
\hline
$z_{photo} < 1.0$   &  $<\delta z>$  &  $\sigma_{\mathrm{MAD}}$  &  $\eta_{>0.15}$  \\
\hline
CLAP  &  0.00079  &  0.01499  &  $0.75\%$  \\
\hline
SCL Prediction  &  -0.00152  &  0.01488   &  $0.74\%$  \\
\hline
Inception Net + Bias Corr. \citep{Lin2022}  &  -0.00375  &  0.01991  &  $2.82\%$  \\
\hline
Inception Net (Treyer et al. in prep.)  &  -0.00323  &  0.01455  &  $0.78\%$  \\
\hline
Le Phare  &  -0.00268 &  0.03271  &  $1.67\%$  \\
\hline
\hline
$z_{photo} > 1.0$   &  $<\delta z>$  &  $\sigma_{\mathrm{MAD}}$  &  $\eta_{>0.15}$  \\
\hline
CLAP  &  0.01509  &  0.03156  &  $7.42\%$  \\
\hline
SCL Prediction  &  0.01221  &  0.03024  &  $7.29\%$  \\
\hline
Inception Net + Bias Corr. \citep{Lin2022}  &  0.04133  &  0.06178  &  $13.96\%$  \\
\hline
Inception Net (Treyer et al. in prep.)  &  0.02861  &  0.02958  &  $6.80\%$  \\
\hline
Le Phare  &  0.06206 &  0.06394  &  $14.60\%$  \\
\hline
\end{tabular}
\caption{Statistics for the point estimates obtained by the methods presented in Sect.~\ref{sec:results} for the CFHTLS dataset. The results with $z_{photo} < 1.0$ (covering roughly 90\% of the full sample) and $z_{photo} > 1.0$ are shown separately.} \label{tab:stats_CFHTLS}
\end{table}

\begin{table}\footnotesize
\centering
\begin{tabular}{l | c c c}
\hline
$z_{photo} < 0.6$   &  $<\delta z>$  &  $\sigma_{\mathrm{MAD}}$  &  $\eta_{>0.15}$  \\
\hline
CLAP  &  0.00011  &  0.01557  &  $0.30\%$  \\
\hline
CLAP (No NIR)  &  0.00006  &  0.01822  &  $0.25\%$  \\
\hline
SCL Prediction  &  -0.00087  &  0.01552  &  $0.32\%$  \\
\hline
Hybrid Net \citep{Li2022GaZNets}  &  -0.00375  &  0.01447  &  $0.25\%$  \\
\hline
VGG Net \citep{Li2022GaZNets}  &  0.00368  &  0.02428  &  $0.35\%$  \\
\hline
MLPQNA  &  0.00142 &  0.02518  &  $1.06\%$  \\
\hline
ANNz2  &  -0.01399 &  0.03436  &  $3.13\%$  \\
\hline
BPZ  &  -0.00327 &  0.03105  &  $3.43\%$  \\
\hline
\hline
$z_{photo} > 0.6$   &  $<\delta z>$  &  $\sigma_{\mathrm{MAD}}$  &  $\eta_{>0.15}$  \\
\hline
CLAP  &  0.01712  &  0.04719  &  $17.95\%$  \\
\hline
CLAP (No NIR)  &  0.02491  &  0.07125  &  $27.32\%$  \\
\hline
SCL Prediction  &  0.01257  &  0.03375  &  $12.80\%$  \\
\hline
Hybrid Net \citep{Li2022GaZNets}  &  0.02149  &  0.02745  &  $9.94\%$  \\
\hline
VGG Net \citep{Li2022GaZNets}  &  0.02741  &  0.06612  &  $25.61\%$  \\
\hline
MLPQNA  &  0.01612 &  0.07637  &  $30.34\%$  \\
\hline
ANNz2  &  -0.05409 &  0.04152  &  $17.97\%$  \\
\hline
BPZ  &  0.02654 &  0.04484  &  $14.33\%$  \\
\hline
\end{tabular}
\caption{Statistics for the point estimates obtained by the methods presented in Sect.~\ref{sec:results} for the KiDS dataset. The results with $z_{photo} < 0.6$ (covering roughly 90\% of the full sample) and $z_{photo} > 0.6$ are shown separately.} \label{tab:stats_KiDS}
\end{table}

\section{Local spatial distribution of the nearest neighbours} \label{sec:spatial_nn}

\begin{figure*}
\centering
\includegraphics[width=1.0\linewidth]{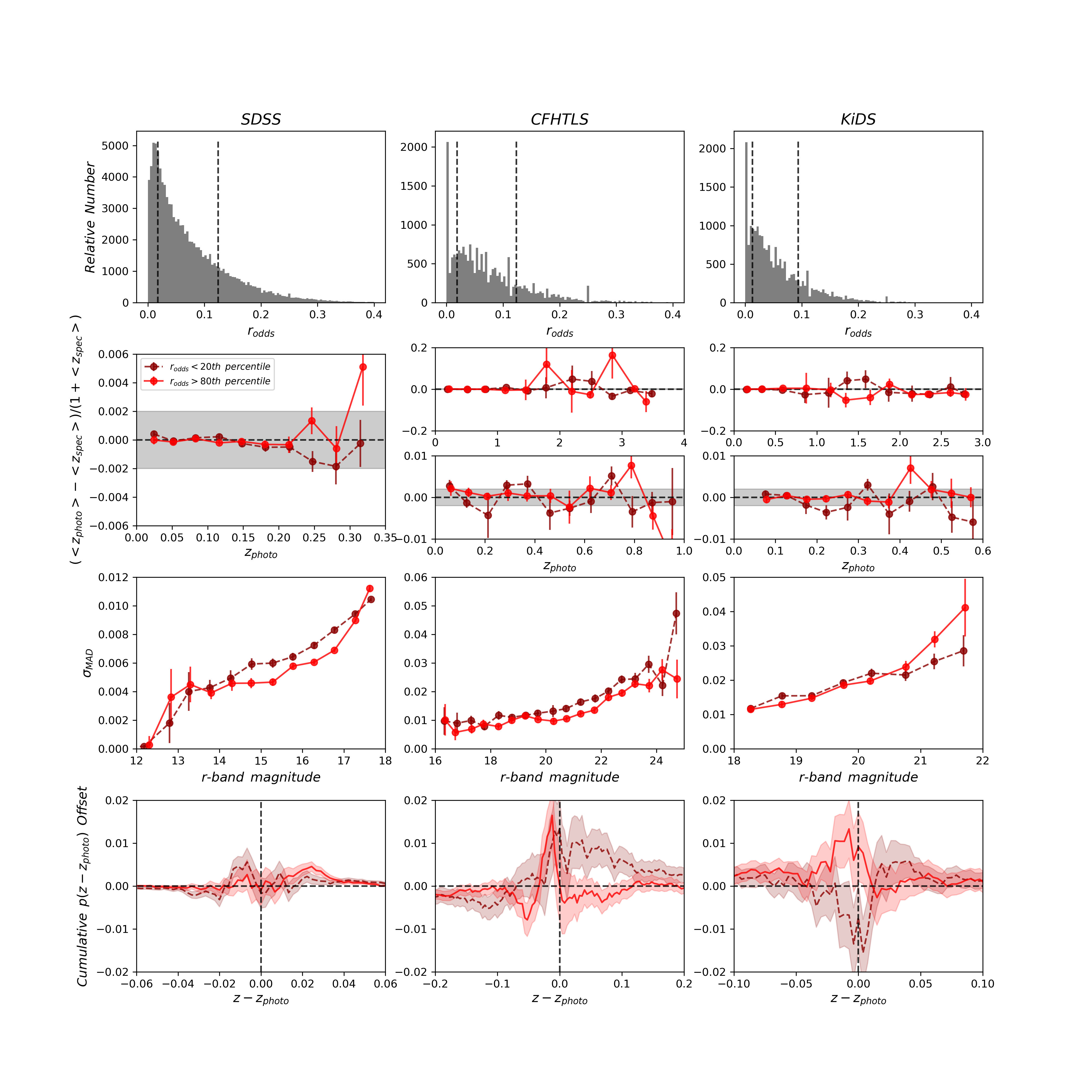}
\caption{Comparisons between two subsamples selected by the odds ratio $r_{odds}$ that characterises the local spatial distribution of the nearest neighbours of each data instance in the latent space (i.e. $r_{odds} < 20th\,\,\,percentile$ and $r_{odds} > 80th\,\,\,percentile$). The results are shown using the SDSS, CFHTLS, and KiDS target samples. For each dataset, only one of the ten CLAP models in the ensemble is selected for demonstration. \textit{First row:} $r_{odds}$ distributions. In each panel, the vertical dashed lines indicate the 20th and 80th percentiles of the distribution. \textit{Second row:} Mean redshift residuals as a function of photometric redshift $z_{photo}$. Same as in Fig.~\ref{fig:main_point_est}, the results for the CFHTLS dataset and the KiDS dataset are shown in the low redshift ranges in addition to the full ranges. \textit{Third row:} $\sigma_{\mathrm{MAD}}$ as a function of $r$-band magnitude. \textit{Fourth row:} Cumulative offsets $\Delta F_{p_{\sum}-h}$ between each of the stacked recentred probability densities $p_{\sum}(z - z_{photo})$ and the corresponding $z_{spec} - z_{photo}$ histogram.}
\label{fig:odds_pointest_pdfsc}
\end{figure*}

\begin{figure*}
\centering
\includegraphics[width=1.0\linewidth]{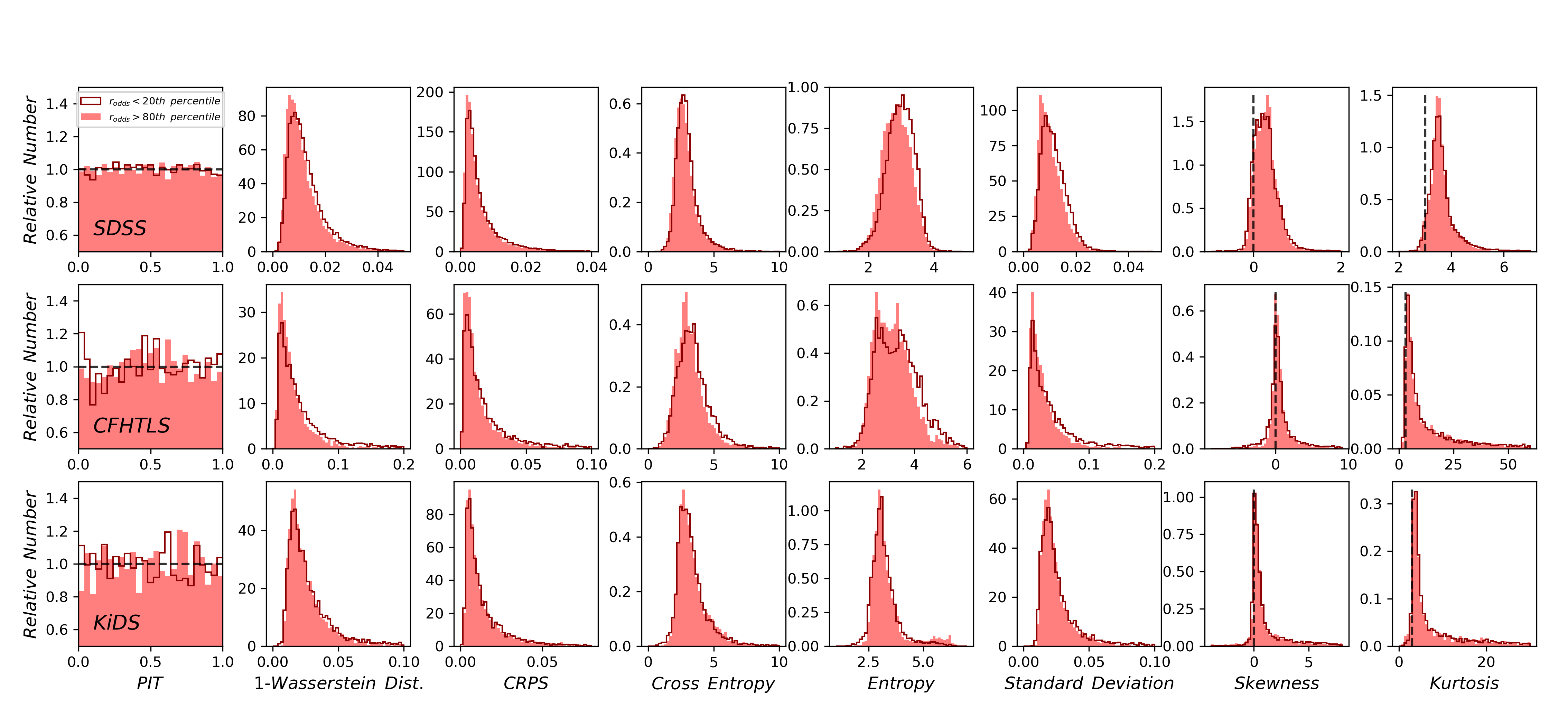}
\caption{Same as Fig.~\ref{fig:main_pdf_prop}, but showing the comparisons between the two subsamples with $r_{odds} < 20th\,\,\,percentile$ and $r_{odds} > 80th\,\,\,percentile$. For each dataset, only one of the ten CLAP models in the ensemble is selected for demonstration.}
\label{fig:odds_pdf_prop}
\end{figure*}

For a finite latent space established by CLAP using a given training sample, how the nearest neighbours are spatially distributed in the local neighbourhood would differ for each query data instance because of uneven data coverage. In particular, the local spatial distributions of the nearest neighbours are likely to be skewed for query data instances close to or beyond the boundary of the established latent space, or located in regions with sparse data coverage, which may lead to biased estimation of probability densities. 

To check this effect, we first leveraged an `odds' ratio to characterise the local spatial distributions of the nearest neighbours, defined as follows. For each data instance, we determine the centre of mass $C$ of its nearest neighbours in the latent space. We define a hyperplane that goes through the projection $O$ of the given data instance and is also perpendicular to $OC$, the link between the two points. This hyperplane divides the local neighbourhood into two halves. Then, the odds ratio is defined as $r_{odds} = N_{\backslash C} / N_C $, where $N_C$ is the number of the nearest neighbours falling in the half with $C$, and $N_{\backslash C}$ is the number of the nearest neighbours in the other half. A lower $r_{odds}$ indicates a more skewed spatial distribution, while the $r_{odds}$ of an even distribution is close to one.

For each of the SDSS, CFHTLS, and KiDS target samples, we compare two subsamples enclosed by two percentiles (i.e. 20th and 80th, respectively) from the overall $r_{odds}$ distribution, using one of the ten CLAP models in the ensemble. These percentiles are selected so that the two subsamples represent the two ends of the $r_{odds}$ distribution and the number of data instances in each subsample is not too low. The results are illustrated in Figs.~\ref{fig:odds_pointest_pdfsc} and \ref{fig:odds_pdf_prop}. We found that the $r_{odds}$ distributions for all the datasets are far from uniformity, with most data populated in the low $r_{odds}$ bins. The shown results such as the cumulative offsets $\Delta F_{p_{\sum}-h}$ and the PIT distributions give an impression that the probability densities from the two subsamples would not be as well calibrated as stated in the main text. However, these results may be simply under the impact of mismatch, since the $r_{odds}$-selected subsamples no longer follow the original redshift-data distribution for the training sample and the target sample. There may also be random errors due to low statistics. Therefore, these results do not conclusively indicate that the estimation of probability densities for either subsample is biased. In order to avoid artificially introducing extra biases, we did not make any selection on $r_{odds}$ in our main experiments. To ensure the robustness of CLAP, future work will need to further investigate the local spatial distributions of the nearest neighbours in the latent space, and focus on evaluating the properties of the sparse regions and the boundary.

\end{appendix}

\end{CJK*}
\end{document}